%% file: main.tex
\documentclass[acmsmall]{acmart}
 \AtBeginDocument{%
   \providecommand\BibTeX{{%
     \normalfont B\kern-0.5em{\scshape i\kern-0.25em b}\kern-0.8em\TeX}}}

\usepackage{siunitx}
\usepackage[flushleft]{threeparttable}
\usepackage{soul}

\begin{document}

\title{Guarding Machine Learning Hardware Against Physical Side-Channel Attacks}

\author{Anuj Dubey}
\authornotemark[1]
\email{aanujdu@ncsu.edu}
\affiliation{%
  \institution{North Carolina State University}
  \streetaddress{891 }
  \city{Raleigh}
  \state{North Carolina}
  \country{USA}
  \postcode{27606}
}

\author{Rosario Cammarota}
\affiliation{%
  \institution{Emerging Security Lab, Intel Corporation}
  \city{San Diego}
  \state{California}
  \country{USA}
  \postcode{92127}
}

\author{Vikram Suresh}
\affiliation{%
  \institution{Circuit Research Lab, Intel Corporation}
  \city{Hillsboro}
  \state{Oregon}
  \country{USA}
  \postcode{97124}
}

\author{Aydin Aysu}
\email{aaysu@ncsu.edu}
\affiliation{%
  \institution{North Carolina State University}
  \streetaddress{891 }
  \city{Raleigh}
  \state{North Carolina}
  \country{USA}
  \postcode{27606}
}


\begin{abstract}
Machine learning (ML) models can be trade secrets due to their development cost. Hence, they need protection against malicious forms of reverse engineering (e.g., in IP piracy). With a growing shift of ML to the edge devices, in part for performance and in part for privacy benefits, the models have become susceptible to the so-called physical side-channel attacks.

ML being a relatively new target compared to cryptography poses the problem of side-channel analysis in a context that lacks published literature. The gap between the burgeoning edge-based ML devices and the research on adequate defenses to provide side-channel security for them thus motivates our study. Our work develops and combines different flavors of side-channel defenses for ML models in the hardware blocks. We propose and optimize the \emph{first defense based on Boolean masking}. We first implement all the masked hardware blocks. We then present an adder optimization to reduce the area and latency overheads. Finally, we couple it with a shuffle-based defense.

We quantify that the area-delay overhead of masking ranges from 5.4× to 4.7× depending on the adder topology used and demonstrate a first-order side-channel security of millions of power traces. 
Additionally, the shuffle countermeasure impedes a straightforward second-order attack on our first-order masked implementation.

\end{abstract}

\setcopyright{acmcopyright}
\acmJournal{JETC}
\acmYear{2021} \acmVolume{1} \acmNumber{1} \acmArticle{1} \acmMonth{1} \acmPrice{15.00}\acmDOI{10.1145/3465377}

\ccsdesc[500]{Security and privacy~Side-channel analysis and countermeasures}

\keywords{side-channel attack, neural networks, masking}

\maketitle

\input{1.Intro}
\input{2.Threat}
\input{3.Back}
\input{4.Defence}
\input{5.Results}
\input{6.Disc}

\section{Conclusions and Future Outlook}
Physical side-channel analysis of neural networks is a new, promising direction in hardware security where the attacks are rapidly evolving compared to defenses.
We proposed the first fully-masked neural network, demonstrated the security with up to 2M traces, and quantified the overheads of a potential countermeasure.
We addressed the key challenge of masking integer addition~\cite{dubey2019maskednet} through Boolean masking. 
We furthermore presented ideas on how to mask the unique linear and non-linear computations of a fully-connected neural network that do not exist in cryptographic applications. We also demonstrated how to couple a first-order masked design with a lightweight shuffle countermeasure to provide additional second-order side-channel security. 
The combination of the two defenses can raise the side-channel resilience to a level that succeeding a theoretical/cryptanalytic attack may need fewer queries.

The large variety in neural network architectures in terms of the quantization-level and the types of layer operations (e.g., Convolution, Maxpool, Softmax), and activation functions (e.g., ReLU, Sigmoid, Tanh) presents a large design space for neural network side-channel defenses. 
This work focused on BNNs, because we feel that BNNs are both excellent candidates for resource-constrained devices deployed at the edge and the closest flavors of ML algorithms to block ciphers (on which a large chunk of the side-channel literature focuses).
Our ideas serve as a benchmark to analyze the vulnerabilities that exist in neural network computations and to construct more robust and efficient countermeasures.

\section{Acknowledgements}
This project is supported in part by NSF under Grants No. 1943245 and SRC GRC Task 2908.001. We also thank Dr. Sohrab Aftabjahani and Dr. Avinash Varna for their valuable feedback on the design.

\bibliographystyle{ACM-Reference-Format}
\bibliography{papers.bib}

\end{document}

%% file: 1.Intro.tex
\section{Introduction}

Machine learning (ML) has become ubiquitous in various domains such as healthcare~\cite{ml-heathcare}, 
automotive~\cite{ml-auto}, and cybersecurity~\cite{ml-cybersec}, among others. 
ML has also made significant advances in terms of performance albeit with increased development costs---e.g., training a recent ML model is estimated to cost over \$4.6M~\cite{costML}.  
The increasing demand for ML models along with their costly development has created a favorable market for selling ML models directly as a service to the customers~\cite{mlaas}. The model provider can either host the trained model on the cloud or deploy it directly on an edge device like a surveillance camera~\cite{aicam}. The recent trend is indeed to directly deploy the ML model on an edge platform for better performance and privacy~\cite{ml2edge}. 
These valuable ML model IPs should be kept confidential. High fidelity model extraction attacks have already been proposed in prior literature that use mathematical and/or cryptanalytic approaches to reverse engineer the internals of the model~\cite{jagielski2019high,carlini2020cryptanalytic}. 
The shift from cloud-based servers to edge devices has additionally exposed the ML models to the easily applicable and extremely potent physical side-channel attacks~\cite{batina2018csi,wei2018know,dubey2019maskednet,yudeepem,emsca-act,xiang2020open,regazzoni2020machine}. Furthermore, the number of queries for success in side-channels attacks is orders of magnitude lower than that of the mathematical model extraction attacks. 
ML models may also be more vulnerable to such attacks when compared with cryptographic implementations because the latter usually has a built-in key-refresh mechanism unlike ML models, which places an upper-bound on the number of traces that can be captured by an adversary~\cite{key-refresh}.

Physical side-channel attacks exploit the information leakage of the secret through physical properties of the device like power-consumption and electromagnetic emanations (EM). The Differential Power Analysis (DPA), for instance, exploits the inherent correlations between the secret-key dependent data being processed and the power consumption of the device \cite{C:KocJafJun99}. Many cryptographic implementations have been shown to be vulnerable to such attacks since then \cite{mangard2008power,TCHES:PSKH18,chen2015differential}. Researchers have accordingly proposed effective ways to mitigate such attacks~\cite{des-dup,ICICS:NikRecRij06,C:IshSahWag03,tiri2004logic,mdrp}. However, the side-channel analysis research focused primarily on protecting cryptographic implementations because it was the only domain that required confidentiality. But lately, ML models have also become lucrative targets for side-channel attacks. Stealing the model internals also assists in adversarial attacks that aim to create misclassifications in the model for malicious purposes \cite{advlearn}.

Physical side-channel attacks are easily applicable if the adversary has physical access to the device, which is why edge-based ML accelerators are unsurprisingly susceptible to such attacks \cite{dubey2019maskednet,wei2018know,batina2018csi,yudeepem}. 
With the recent developments in remote physical side-channel attacks, it is also possible to extend reverse engineering of ML models to a multi-tenant cloud-based FPGA setting~\cite{zhao2018fpga,insidejob,nn-remote}. However, the research on developing adequate countermeasures is still quite immature. With the latest market research predicting a tremendous growth in the sales of edge-based ML hardware in the coming years~\cite{edge-ai}, there is an urgent need to develop efficient and robust side-channel defenses for ML applications.

The existing works on building countermeasures against power-based side-channel attacks on ML accelerators extend the ones developed for the cryptographic applications. 
The only two available techniques from our earlier works either utilize a combination of masking and hiding for side-channel resilience or a full Boolean masking approach \cite{dubey2019maskednet,bomanet}. 
Hiding aims to equalize power consumption throughout the execution typically using a precharge and differential logic based circuit. 
Masking splits (or encodes) the original secret into multiple statistically independent shares to break the correlation of the secret data with the power consumption. 
The former approach is cost-conscious but challenging to implement effectively due to the precise control required in back-end flow for the hiding countermeasure. 
The latter approach of using Boolean masking is relatively easy to implement effectively because it is an algorithm-level defense and does not require a precise back-end control. 
However, it incurs higher performance and area overheads when implemented trivially.

In this article, we extend our previous work on Boolean masking that was published at the IEEE/ACM International Conference on Computer Aided Design 2020 \cite{bomanet}. 
The earlier work presents the first fully masked ML accelerator design that provides protection against power side-channel attacks.
The extensions over the previous work include the following.

\begin{itemize}
    \item Our previous work used a pipelined ripple-carry architecture for the masked adder design which had a latency of 100 cycles. In the extension, we design and implement a masked Kogge-Stone architecture for the adder that reduces its latency from 100 cycles to 31 cycles. We quantify that such a change improves the area-delay overheads of the whole design from 5.4$\times$ to 4.7$\times$ without changing the side-channel security. We quantify the area reduction mainly in terms of the number of LUTs because LUTs are the limiting factor in FPGAs; flip flops are available in abundance\footnote{The target FPGA is Spartan-6 in which each slice consists of 4 LUTs but 8 flip flops.}.
    Section \ref{sec:ksa} discusses the details of the design, Section \ref{sec:sca-ksa} illustrates the side-channel resilience, and Section \ref{sec:maskov} presents the implementation results.
    
    \item We propose a shuffle countermeasure to improve the side-channel resilience and make it more difficult to conduct a second-order side-channel attack on our first-order masked design.
    Specifically, the scheme randomize the sequence in which neurons are computed for each layer by adopting the random start index (RSI) method originally developed for protecting AES implementations~\cite{rsi1,shuffle-survey}. Shuffling introduces noise in the temporal domain and increases the number of measurements for a successful attack by $N^2$, where $N$ is the number of hidden layer nodes. We discuss the details of the implementation in Section \ref{sec:shuffle}. We also demonstrate that shuffling helps to increase even the second-order side-channel security of the design up to at least 3M traces in a low noise setup.
    
    \item We also explore an alternative implementation where we switch our earlier proposed adder using the pipelined Trichina's AND gate~\cite{TKL05} with a masked look-up table (LUT) having 5 inputs. 
    We show that such a design further reduces the area-delay product by 2$\times$ but does leak information against a first-order side-channel attack.
    The leakage, however, is relatively smaller compared to an unprotected design with leakages becoming statistically significant only after capturing 270k measurements. 
    We discuss the details of the implementation and quantify the side-channel leakage in Section \ref{sec:luttrichina}.
\end{itemize}

The rest of this manuscript is organized as follows: Section 2 discusses our assumptions on the adversary and the victim; Section 3 briefly discusses related work on ML model extraction, existing side-channel defenses in cryptography, neural networks and describes our baseline hardware design; Section 4 describes in detail the hardware design of our proposed masked neural network components; Section 5 discusses the implementation of the shuffle countermeasure; Section 6 presents our results on hardware implementation results and the side-channel evaluation; Section 7 discusses potential ways to reduce the costs further or provide provable security; Finally, we conclude our paper in Section 8. 

%% file: 2.Threat.tex
\section{Threat Model}
We adopt the standard DPA threat model in which an adversary has direct physical access to the target device running inference~\cite{batina2018csi,dubey2019maskednet,kocher2011introduction}, or can obtain power measurements remotely when the device executes neural network computations~\cite{zhao2018fpga}. The adversary can control the inputs and observe the corresponding outputs from the device as in chosen-plaintext attacks.  
\begin{figure}[t!]
    \includegraphics[scale=0.8]{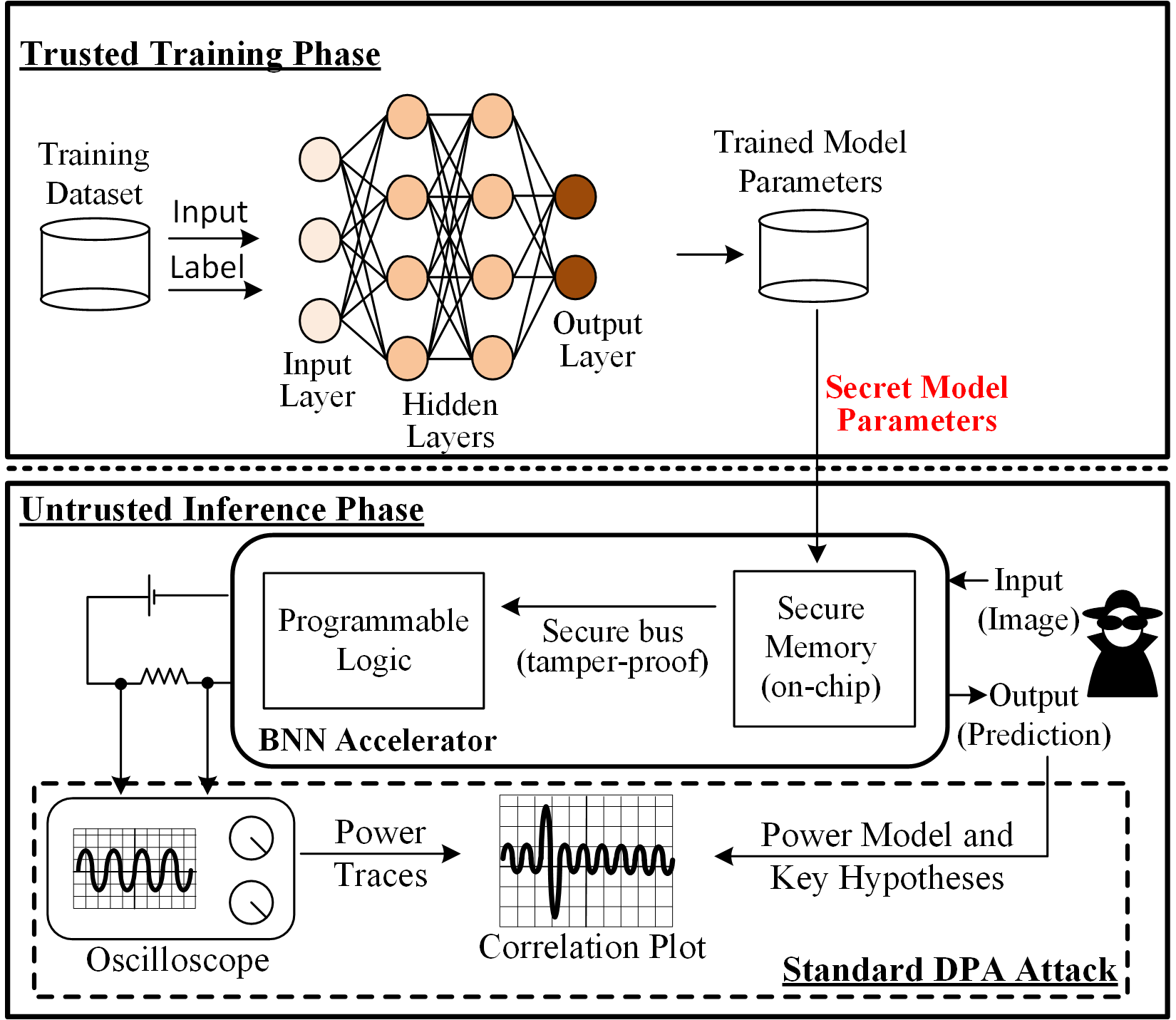}
    \caption{Standard DPA threat model applied to ML model stealing where the trained neural network is deployed on an edge device running in an untrusted environment.}
    \label{fig:threat}
    \vspace{-2em}
\end{figure}

Fig. \ref{fig:threat} shows our threat model where the training phase is trusted but the trained model is deployed to an inference engine operating in an untrusted environment. The adversary is after the trained model parameters (e.g., weights and biases)---input data privacy is out of scope \cite{wei2018know}.

We assume that the trained ML model is stored in a protected memory and the standard techniques are used to securely transfer it (i.e., bus snooping attacks are out of scope) \cite{mlIPprot}.
The adversary, therefore, has gray-box access to the device, i.e., it knows all the design details up to the level of each logic gate but does not know the trained ML model. 
We restrict the secret variables to just the parameters and not the hyperparameters such as the number of neurons, following earlier work~\cite{juuti2019prada,tramer2016stealing,dubey2019maskednet}. 
In fact, an adversary will still not be able to clone the model with just the hyperparameters if it does not possess the required compute power or training dataset. This is analogous to the scenario in cryptography where an adversary, even after knowing the implementation of a cipher, cannot break it without the correct key.

We target a hardware implementation of the neural network, not software. The design fully fits on the FPGA. Therefore, it does not involve any off-chip memory access and executes with constant-flow in constant time. These attributes make the design resilient to any type of digital (memory, timing, access-pattern, etc.) side-channel attack. However, the physical side-channels like power and EM emanations still exist; we address the power-based side-channel leakages in our work. 
Other implementation attacks on neural networks such as the fault attacks are out of scope~\cite{breier2018practical,breier2020sniff}.

%% file: 3.Back.tex
\section{Background and Related Work}
This section presents related work on the privacy of ML applications, the current state of side-channel defenses, preliminaries on BNNs, and our BNN hardware design.
\subsection{ML Model Extraction}
Recent developments in the field of ML point to several motivating scenarios that  demand asset confidentiality. 
Firstly, training is a computationally-intensive process and hence requires the model provider to invest money in high-performance compute resources (e.g., a GPU cluster). The model provider might also need to invest money to buy a labeled dataset for training or label an unstructured dataset. Therefore, knowledge about either the parameters or hyperparameters can provide an unfair business advantage to the user of the model, which is why the ML model should be private. 

Theoretical model extraction analyzes the query-response pair obtained by repeatedly querying an unknown ML model to steal the parameters 
\cite{jagielski2019high,oh2019towards,RST19,carlini2020cryptanalytic}. 
This type of attack is similar to the class of theoretical cryptanalysis in the cryptography literature. 
Digital side-channels, by contrast, exploit the leakage of secret-data dependent \emph{intermediate computations} like access-patterns or timing in the neural network computations to steal the parameters \cite{yan2018cache,duddu2018stealing,dong2019floating,hua2018reverse,hu2019neural}, which can usually be mitigated by making the secret computations constant-flow and constant-time. Physical side-channels target the leak in the physical properties like CMOS power-draw or electromagnetic emanations that will still exist in a constant-flow/constant-time algorithm's implementation~\cite{batina2018csi,wei2018know,dubey2019maskednet,yudeepem,emsca-act,xiang2020open}. Mitigating physical side-channels are thus harder than digital side-channels in hardware accelerator design and has been extensively studied in the cryptography community.

\subsection{Side-Channel Defenses}
The researchers have proposed numerous countermeasures against DPA. These countermeasures can be broadly classified as either \emph{hiding-based} or \emph{masking-based}. The former aims to make the power-consumption constant throughout the computation by using power-balancing techniques \cite{tiri2004logic,nassar2010bcdl,yu2007secure}. The latter splits the sensitive variable into multiple statistically independent shares to ensure that the power consumption is independent of the sensitive variable throughout the computation \cite{CHES:AkkGir01,CHES:GolTym02,BGK05,OMPR05,TKL05,C:IshSahWag03}. 

The security provided by hiding-based schemes hinges upon the precision of the back-end design tools to create a near-perfect power-equalized circuit by balancing the load capacitances and synchronizing their activity across the leakage prone paths. This is not a trivial task and prior literature shows how a well-balanced dual-rail based defense is still vulnerable to localized EM attacks \cite{immler2017your}. By contrast, masking transforms the algorithm itself to work in a secure way by never evaluating the secret variables directly. 
This keeps the security largely independent of the back-end design and makes masking a favorable choice over hiding.

\subsection{Neural Network Classifiers}
Neural network algorithms learn how to perform a certain task. In the learning phase, the user sends a set of inputs and expected outputs to the machine (a.k.a., training), which helps it to approximate (or learn) the function mapping the input-output pairs. The learned function can then be used by the machine to generate outputs for unknown inputs (a.k.a., inference).

A neural network consists of units called neurons (or nodes) and these neurons are usually grouped into layers. The neurons in each layer can be connected to the neurons in the previous and next layers. Each connection has a weight associated with it, which is computed during the training. The neurons work in a feed-forward fashion passing information from one layer to the next. 

The weights and biases can be initialized to be random values or a carefully chosen set before training \cite{tlearning}. 
These weights and biases are the \emph{critical parameters} that our countermeasure aims to protect.
During training, a set of inputs along with the corresponding labels are fed to the network. The network computes the error between the actual outputs and the labels and tunes the weights and biases to reduce it, converging to a state where the accuracy is acceptable. 

\subsection{Binarized Neural Networks}
\begin{figure}
    \vspace{-1em}
    \includegraphics[scale=0.9]{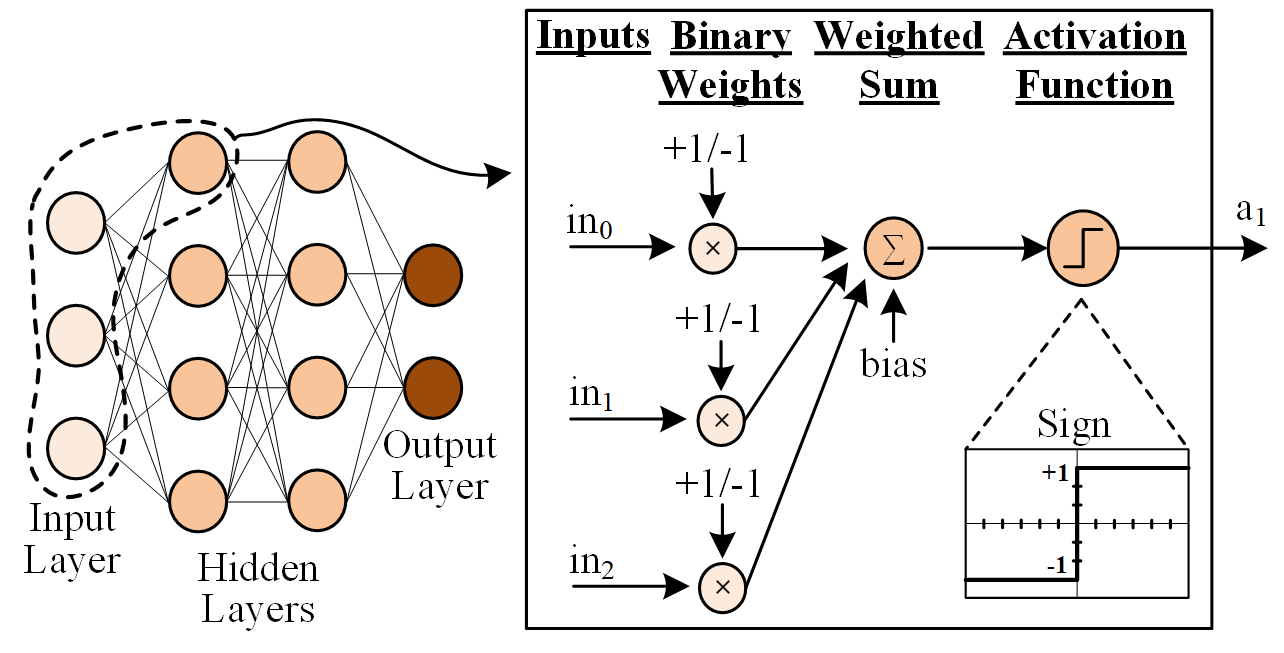}
    \vspace{-1em}
    \caption{A typical Binarized Neural Network where the neuron performs weighted summations of the input pixels with binarized weights, and the activation function is a sign function.}
    \vspace{-1em}
    \label{fig:neuron}
\end{figure}

Fig.~\ref{fig:neuron} depicts the neuron computation in a fully-connected BNN. The weights and biases of a neural network are typically floating-point numbers. However, high area, storage costs, and power demands of floating-point hardware do not fare well with the requirements of the resource-constrained edge devices. Binarized Neural Networks (BNNs) \cite{courbariaux2016binarized}, with their low hardware cost and power needs fit very well in this use-case while providing a reasonable accuracy. BNNs restrict the weights and activations to binary values (+1 and -1), which can easily be represented in hardware by a single bit. This reduces the storage costs for the weights from floating-point values to binary values. The XNOR-POPCOUNT operation implemented using XNOR gates replaces the large floating-point multipliers resulting in a huge area and performance gain \cite{rastegari2016xnor}. In Fig.~\ref{fig:neuron}, the neuron in the first hidden layer multiplies the input values with their respective binarized weights. The generated products are added to the bias, and the result is fed to the activation function, which is a sign function that binarizes the non-negative and negative inputs to +1 to -1, respectively. Hence, the activations in the subsequent layer are also binarized. 

Prior works have demonstrated the computational efficiency of XNOR-POPCOUNT based arithmetic operations in DNNs. A prior proposed GPU kernel exploiting the XNOR-POPCOUNT operations achieved $23\times$ faster matrix multiplication compared to a naive baseline implementation~\mbox{\cite{Courbariaux2016BinaryNetTD}}. The designed kernel is $3.4\times$ faster than cuBLAS and the MLP runs $7\times $ faster using the XNOR-POPCOUNT kernel compared to the baseline. Well known semiconductor companies like Xilinx, Intel, and Apple are showing interest in BNNs due to these advantages~\mbox{\cite{umuroglu2017finn, knag2020617, apple-xnor}}. Owing to their low memory footprint and lightweight nature of operations, BNNs are considered attractive for edge applications such as FPGA-accelerators~\mbox{\cite{cheng2019towards}}, cryptographic neural network inference systems~\mbox{\cite{xiaoning2020leia}}, and for designing low-bitwidth ConvNets~\mbox{\cite{zhou2016dorefa}}, among many other applications. Despite their low-bitwidth operations, the accuracy obtained by BNNs is comparable to that obtained by full-precision neural networks. For instance, the accuracy loss of a BNN-ConvNet with 1-bit weights and activations was found to be less than $0.5\%$ compared to a full-precision ConvNet with seven convolutional layers and one dense layer~\mbox{\cite{zhou2016dorefa}} when evaluated on the Google Street View House Number (SVHN) dataset. Similarly, another work~\mbox{\cite{lin2017towards}} achieved less than $5\%$ accuracy loss on the ImageNet dataset~\mbox{\cite{deng2009imagenet}}, a challenging dataset notorious for its complexity in the computer vision community, using a ResNet-like network built entirely using binary convolution blocks, compared to a full precision network.


\subsection{Our Baseline BNN Hardware Design}
We consider a BNN having an input layer of 784 nodes, 3 hidden layers of 1010\footnote{The number of hidden layer nodes can change based on the type of adder used, which is explained later in the manuscript.} nodes each, and an output layer of 10 nodes. The 784 input nodes denote the 784 pixel values in the 28$\times$28 grayscale images of the Modified National Institute of Standards and Technology (MNIST) database and 10 output nodes represent the 10 output classes of the handwritten numerical digit.
\begin{figure}
    \includegraphics{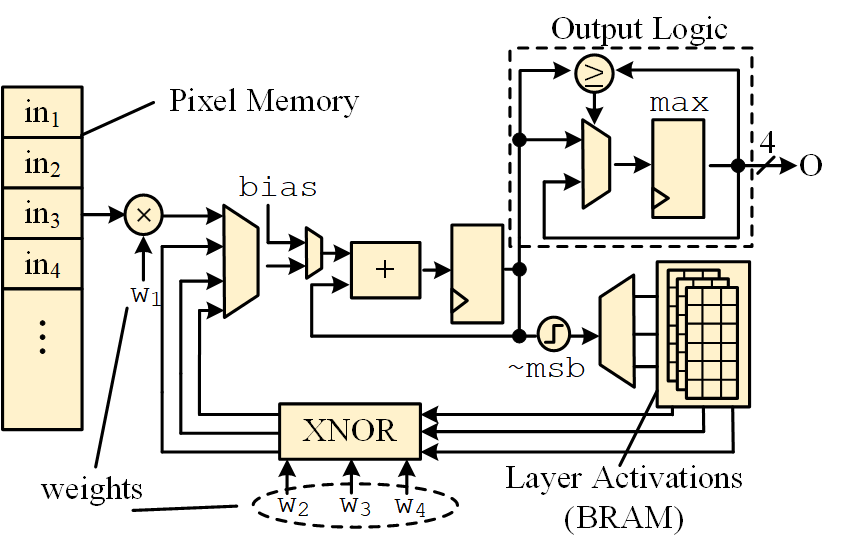}
    \caption{The hardware design of baseline unmasked BNN. The hardware performs weighted summations using a single adder in the input layer and using XNOR-POPCOUNT logic in the hidden layers. The sign function is implemented by simply inverting the MSB of the weighted summation. The output logic finds the maximum weighted summation out of the 10 output layer nodes sequentially.}
    \label{fig:base-arch}
\end{figure}

\subsubsection{Weighted Summations}
We choose to use a single adder in the design and sequentialize all the additions in the algorithm to reduce the area costs. Fig.~\ref{fig:base-arch} shows our baseline BNN design. The computation starts from the input layer pixel values stored in the Pixel Memory. For each node of the first hidden layer, the hardware multiplies 784 input pixel values one by one and accumulates the sum of these products. The final summation is added with the bias reusing the adder with a multiplexed input and fed to the activation function. 
The hardware uses XNOR and POPCOUNT\footnote{The POPCOUNT operation also involves an additional step of subtracting the number of nodes (1010) from the final sum, which can be done as part of bias addition step.} operations to perform weighted summations in the hidden layers. The final layer summations are sent to the output logic. 

\begin{figure}
  \centering
    \includegraphics[width=0.35\textwidth]{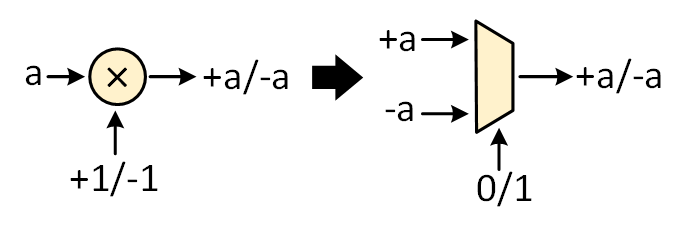}
    \caption{The multiplication of input pixel $a$ with the weight (+1/-1) is implemented as a 9-bit 2$\times$1 multiplexer in the baseline BNN design.}
    \vspace{-1em}
    \label{fig:mmul}
\end{figure}

In the input layer computations, the hardware multiplies an 8-bit unsigned input pixel value with its corresponding weight. The weight values are binarized to either 0 or 1 (representing a -1 or +1, respectively). Fig.~\ref{fig:mmul} shows the realization of this multiplication with a multiplexer that takes in the pixel value ($a$) and its 2's complement ($-a$) as the data inputs and weight ($\pm$1) as the select line. The 8-bit unsigned pixel value, when multiplied by $\pm$1, needs to be sign-extended to 9-bits, resulting in a 9-bit 2$\times$1 multiplexer. 

\subsubsection{Activation Function}
\label{sec:actfn}
The activation function binarizes the non-negative and negative inputs to +1 and -1 respectively for each node of the hidden layer. In hardware, this is implemented using a simple NOT gate that takes the MSB of the summations as its input.

\subsubsection{Output Layer}
The summations in the output layer represent the confidence score of each output class for the provided image. Therefore, the final classification result is the class having the maximum confidence score. Fig.~\ref{fig:base-arch} shows the hardware for computing the classification result. As the adder generates output layer summations, they are sent to the output logic block that performs a rolling update of the max register ($max$) if the newly received sum is greater than the previously computed max. In parallel, the hardware also stores the index of the current max node. The index stored after the final update is sent out as the final output of the neural network. The hardware takes 2.8M cycles to finish one inference.

\paragraph{Exemplary DPA Attack on Baseline Implementation.} A DPA adversary can target the 20-bit accumulator register at the output of the adder in Fig. 3. The register accumulates the sum of the products of image pixels and weights in every cycle. Thus, the power consumption of any of those cycles can be modeled by using a hamming distance (HD) model. For example, if the adversary targets the 4$^\text{th}$ cycle, the power model $M$ is given as follows.

\[M = HD ((w_0\times p_0+w_1\times p_1+w_2\times p_2),(w_0\times p_0+w_1\times p_1+w_2\times p_2+w_3\times p_3))\]

where $p_i$ and $w_i$ denote the i$^\text{th}$ image pixel and weight, respectively. The number of hypotheses in this case is 16 (2$^\text{4}$) since the adversary attacks four weights $w_0,w_1,w_2$, and $w_3$.
Since the hardware adds the bias in the 785$^\text{th}$ cycle, the adversary needs to extract the 784 weights first and then construct the power model for the sum with bias to attack it. 

%% file: 4.Defence.tex
\section{Fully Masking the Neural Network}
This section discusses the hardware design and implementation of all components in the masked neural network. 
Prior work on masking of neural networks shows that arithmetic masking alone cannot mask integer addition due to a leakage in the sign-bit \cite{dubey2019maskednet}. Hence, we apply gate-level \emph{Boolean} masking to perform integer addition in a secure fashion. We express the entire computation of the neural network as a sequence of AND and XOR operations and apply gate-level masking on the resulting expression. XORs, being linear, do not require any additional masking, and AND gates are replaced with secure, Trichina style AND gates \cite{TKL05}. Furthermore, we design specialized circuits for BNN's unique components like Masked Multiplexer and Masked Output Layer. 
\begin{figure}
  \centering
    \includegraphics{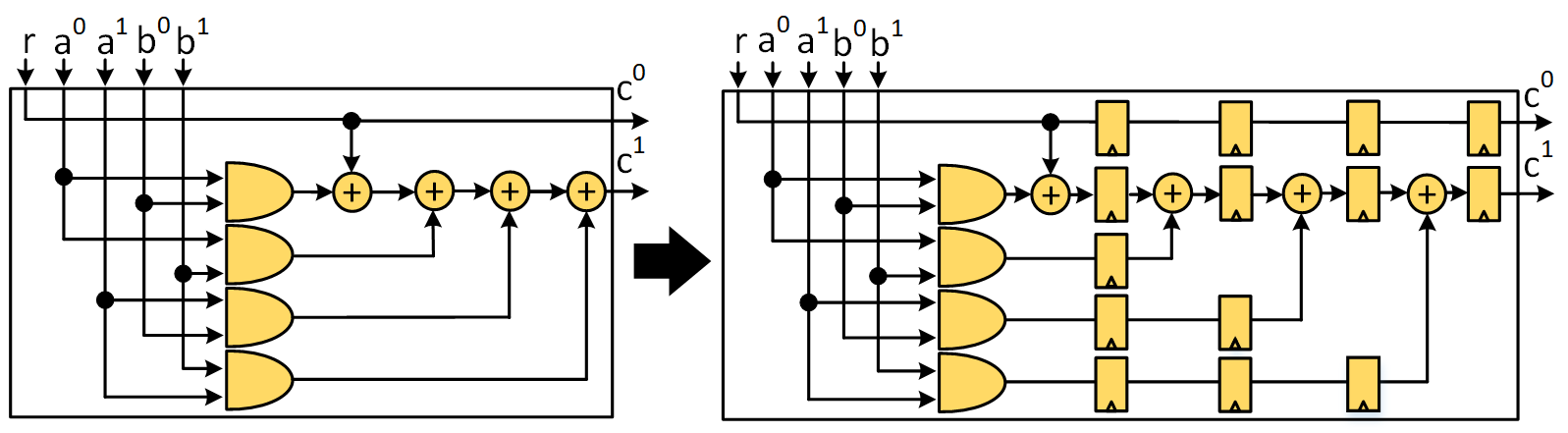}
      \caption{Trichina's AND Gate implementation: glitch-prone (left) and glitch-resistant (right). Flip-flops synchronize arrival of asynchronous gate outputs at XOR gates' inputs to reduce glitches.}
      \label{fig:trichina}
\end{figure}

We first explain the notations in equations and figures. 
Any variable without a subscript or superscript represents an N-bit number. We use the subscript to refer to a single bit of the N-bit number. For example, $a_7$ refers to the $8^{th}$ bit of $a$. 
The superscript in masking refers to the different secret shares of a variable. 
To refer to a particular share of a particular bit of an N-bit number, we use both the subscript and the superscript. For example, $a_{4}^{1}$ refers to the second Boolean share of the $5^{th}$ bit of $a$. If a variable only has the superscript (say $i$), we are referring to its full N-bit $i^{th}$ Boolean share; N can also be equal to 1, in which case $a$ is simply a bit. r (or r${_i}$) denotes a fresh, random bit. The operation $\oplus$ represents a bitwise XOR of operands.

\subsection{A Glitch-Resilient Trichina's AND Gate}
\label{ss:tgate}
Among the closely related masking styles~\cite{reparaz16-2}, we chose to implement Trichina's method due to its simplicity and implementation efficiency. 
Fig.~\ref{fig:trichina} (left) shows the basic structure and functionality of the Trichina's gate, which implements a 2-bit, masked, AND operation of $c=a \cdot b$.
Each input ($a$ and $b$) is split into two shares ($a^0$ and $a^1$ s.t. $a=a^0\oplus a^1$, and $b^0$ and $b^1$ s.t. $b=b^0\oplus b^1$). These shares are sequentially processed with a chain of AND gates initiated with a fresh random bit ($r$).
A single AND operation thus uses 3 random bits.
The technique ensures that output is the Boolean masked output of the original AND function, i.e., $c=c^0 \oplus c^1$, while all the intermediate computations are randomized.

Unfortunately, the straightforward adoption of Trichina's AND gate can lead to information leakage due to glitches~\cite{ICICS:NikRecRij06}. For instance, in Fig.~\ref{fig:trichina} (left) if the products $a_0\cdot b_0$ and $a_0\cdot b_1$ reach the input of second XOR gate before random mask $r$ reaches the input of first XOR gate, the output at the XOR gate will evaluate (glitch) to $(a_0\cdot b_0)\oplus (a_0\cdot b_1)=a_0\cdot(b_0\oplus b_1)$ temporarily, which leads to secret value $b$ being unmasked. 
Therefore, we opted for an extension of the Trichina's AND gate by adding flip-flops to synchronise the arrival of inputs at the XOR gates (see Fig.~\ref{fig:trichina} right). The only XOR gate not having a flip-flop at its input is the leftmost XOR gate in the path of $c_1$, which is not a problem because a glitching output at this gate does not combine two shares of the same variable. 
Similar techniques have been used in past \cite{regGlitch}. Masking styles like the Threshold gates~\cite{CHES:ManPraOsw05, maskedcmosleak,SAC:TirSch06} may be considered for even stronger security guarantees, but they will add further area-performance-randomness overhead.

\begin{figure}
  \centering
    \includegraphics[width=1.0\textwidth]{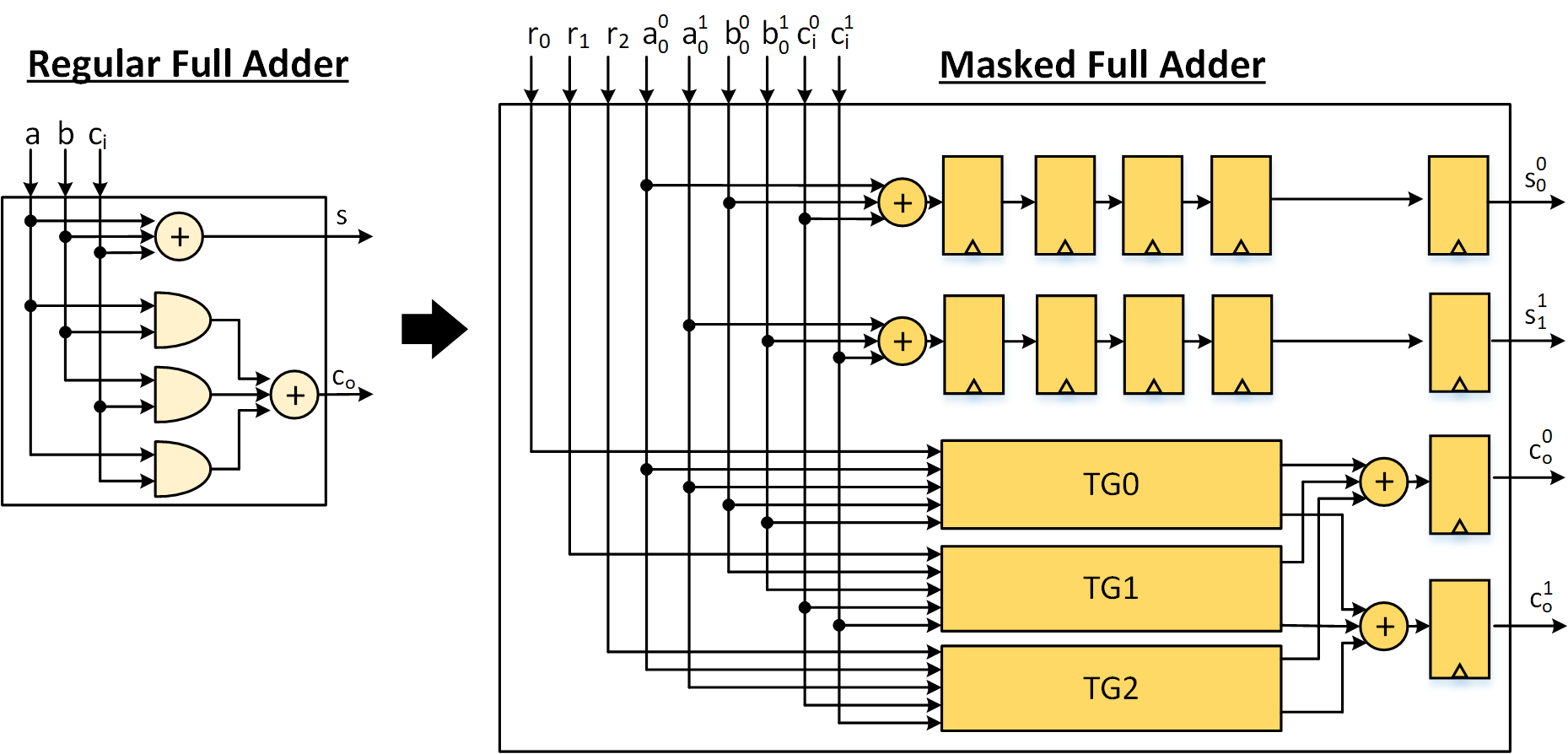}
      \caption{An unmasked Full Adder design (left) and its gate-level masking using Trichina AND Gates (right) that has a latency of 5 cycles.}
      \vspace{-2em}
    \label{fig:bomanet-adders}
\end{figure}

\subsection{Masked Ripple Carry Adder}
We adopt the ripple-carry style of implementation for the adder first. It is composed of N 1-bit full adders where the carry-out from each adder is the carry-in for the next adder in the chain, starting from LSB. Therefore, ripple-carry configuration eases parameterization and modular design of the Boolean masked adders.

\subsubsection{Design of a Masked Full Adder}
A 1-bit full adder takes as input two operands and a carry-in and generates the sum and the carry, which are a function of the two operands and the carry-in. If the input operand bits are denoted by $a$ and $b$ and carry-in bit by $c$, then the Boolean equation of the sum $S$ and the carry $C$ can be described as follows:
\begin{equation}\label{eq:sum}
    S = a\oplus b\oplus c
\end{equation}
\begin{equation}\label{eq:carry}
    C = a\cdot b\;|\;b\cdot c\;|\;c\cdot a
\end{equation}

However, the non-linear OR operation in the carry function is usually replaced with the linear XOR operator to simplify the masking of carry.

\begin{equation}\label{eq:carry}
    C = a\cdot b\oplus b\cdot c\oplus c\cdot a
\end{equation}

Fig.~\ref{fig:bomanet-adders}  shows the regular, 1-bit full adder (on the left), and the resulting masked adder with Trichina's AND gates (on the right). In the rest of the subsection, we will discuss the derivation of the masked full adder equations.

First step is to split the secret variables ($a$, $b$ and $c$) into Boolean shares. 
The hardware samples a fresh, random mask from a uniform distribution and performs XOR with the original variable. If we represent the random masks as $a^{0}$, $b^{0}$ and $c^{0}$, then the masked values $a^{1}$, $b^{1}$ and $c^{1}$ can be generated as follows:
\begin{equation}\label{eq:ma}
    a^1=a\oplus a^0,\;b^1=b\oplus b^0,\;c^1=c\oplus c^0
\end{equation}
A masking scheme always works on the two shares independently without combining them at any point in the operation because that will reconstruct the secret and create a side-channel leak. 

The function of sum-generation is linear, making it easy to directly and independently compute the Boolean shares of $S$:
\begin{equation*}
    S = S^0 \oplus S^1
\end{equation*}

\noindent{where,}
\begin{equation*}
    S^0=a^0\oplus b^0 \oplus c^0,\;S^1=a^1\oplus b^1 \oplus c^1
\end{equation*}

Unlike the sum-generation, carry-generation is a non-linear operation due to the presence of an AND operator. Hence, the hardware cannot directly and independently compute the Boolean shares $C^0$ and $C^1$ of $C$. 
We use the Trichina's construction explained in subsection \ref{ss:tgate} to mask carry-generation.

The hardware uses three Trichina's AND gates to mask the three AND operations in equation (\ref{eq:carry}) using three random masks. This generates two Boolean shares from each Trichina AND operation. At this point, the expression is linear again, and therefore, the hardware can redistribute the terms, similar to the masking of sum operation.
\begin{figure}
  \centering
    \includegraphics[scale=0.9]{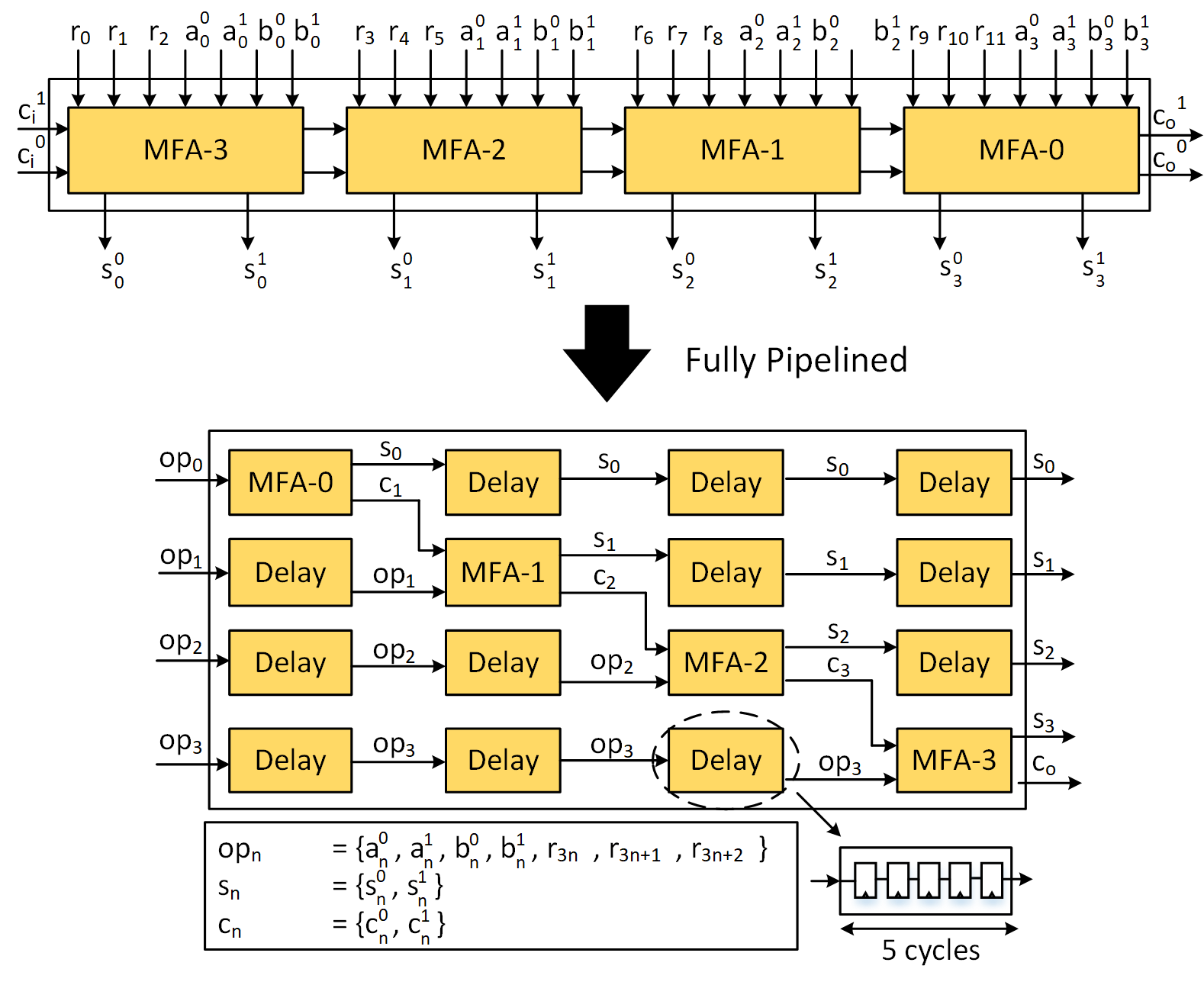}
      \caption{Modular design of a masked 4-bit adder using Masked Full Adders (top) and its pipelined version (bottom). Pipelining ensures a throughput of 1 summation per cycle.}
      \vspace{-2em}
      \label{fig:modularAdd}
\end{figure}
In the following equations, we use $TG(x,y,r)$ to represent the product $x\cdot y$ implemented via Trichina's AND Gate as illustrated in the following equation:
\begin{equation*}
    x\cdot y=TG(x,y,r) = m^0\oplus m^1
\end{equation*}
where $m^0$ and $m^1$ are the two Boolean shares of the product.
Replacing each AND operation in equation (\ref{eq:carry}) with TG, we can write
\begin{equation}\label{eq:tg0}
    TG(a,b,r_0) = d^0\oplus d^1
\end{equation}
\begin{equation}\label{eq:tg1}
    TG(b,c,r_1) = e^0\oplus e^1
\end{equation}
\begin{equation}\label{eq:tg2}
    TG(c,a,r_2) = f^0\oplus f^1
\end{equation}
where $d^0$, $d^1$, $e^0$, $e^1$, $f^0$, and $f^1$ are the output shares from each Trichina Gate. From equations (\ref{eq:carry}), (\ref{eq:tg0}), (\ref{eq:tg1}), and (\ref{eq:tg2}) we get
\begin{equation*}
    carryout = TG(a,b,r_0) \oplus TG(b,c,r_1) \oplus TG(c,a,r_2)
\end{equation*}
Replacing the TGs from equation (\ref{eq:tg0}), (\ref{eq:tg1}), and (\ref{eq:tg2}) and rearranging the terms, we get
\begin{equation*}
    carryout = (d^0\oplus e^0 \oplus f^0) \oplus (d^1 \oplus e^1 \oplus f^1)
\end{equation*}
which can also be written as a combination of two Boolean shares $C^0$ and $C^1$ where


\begin{equation*}
    C^0 = d^0\oplus e^0 \oplus f^0,\;C^1 = d^1 \oplus e^1 \oplus f^1
\end{equation*}
Therefore, we create a masked full adder that takes in the Boolean shares of the two bits to be added along with a carry-in and gives out the Boolean shares of the sum and carry-out. 
\subsubsection{The Modular Design of Pipelined N-bit Full Adder}
The masked full adders can be chained together to create an N-bit masked adder that can add two masked N-bit numbers. Fig.~\ref{fig:modularAdd} (top) shows how to construct a 4-bit masked adder as an example. We pipeline the N-bit adder to yield a throughput of one by adding registers between the full-adders corresponding to each bit (see Fig.~\ref{fig:modularAdd} (bottom)).

\subsection{Masked Kogge Stone Adder}
\label{sec:ksa}
We also propose and implement the masked design for a baseline Kogge Stone adder (KSA) architecture. KSA is a type of parallel prefix adder that has a (logarithmically) lower latency compared to that of the ripple carry adder. KSA builds on the concept of carry look ahead. It starts by computing the \emph{generate} ($g_i$) and \emph{propagate} ($p_i$) bits for each position given by the following equations.

\begin{equation}
\label{eq:gen}
  g_i = a_i\cdot b_i
\end{equation}
\begin{equation}
\label{eq:prop}
 p_i = a_i \oplus b_i 
\end{equation}

\noindent, where $a_i$ and $b_i$ are the $i^{th}$ bits of the operands. The generate bit being 1 implies that the carry will definitely be asserted at that position. The propagate bit being 1 implies that the carry will only be asserted if there is an incoming carry from the previous step. Therefore, the carry can only be asserted if the generate bit is 1 or the propagate bit is 1 with an incoming carry from the previous position. 

This concept of generates and propagates for a single position can be extended to compute the so-called \emph{group generate} and \emph{group propagate} bits that denote if a group of bits do generate or propagate a carry. For example, the following two equations illustrate how to compute the group generate ($G_{1:0}$) and propagate ($P_{1:0}$) for the group of least significant two bits from the individual generates and propagates $g_0,p_0,g_1$ and $p_1$.
\begin{equation*}
    G_{1:0}=g_1\oplus p_1\cdot g_0
\end{equation*}
\begin{equation*}
    P_{1:0}=p_1\cdot p_0
\end{equation*}

Larger groups can be created from smaller groups by combining them in a similar fashion. Let $G_{i:k}$, $P_{i:k}$, $G_{k:j}$ and $P_{k:j}$ represent the group generate and propagate bits for the group of bits from $i^{th}$ to $k^{th}$ position, and $k^{th}$ to $j^{th}$ position, respectively. The generic equation to combine these quantities and get a group generate and propagate for the combined group from $i^{th}$ to $j^{th}$ bit is given below.

\begin{equation}
\label{eq:Ggen}
    G_{i:j}=G_{i:k}\oplus P_{i:k}\cdot G_{k:j}
\end{equation}
\begin{equation}
\label{eq:Gprop}
    P_{i:j}=P_{i:k}\cdot P_{k:j}
\end{equation}

\begin{figure}
  \centering
    \includegraphics{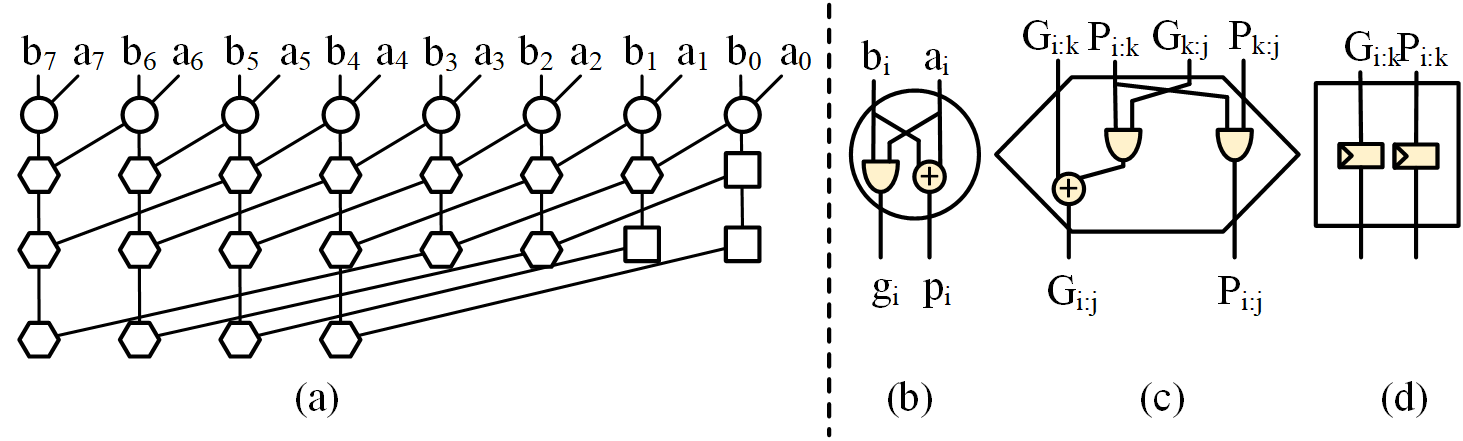}
      \caption{Part (a) shows the topology of an 8-bit KSA and the location of the three different constituents in the design using circles, hexagons and squares. Parts (b), (c), and (d) describe the internals of these 3 constituents, respectively.}
      \label{fig:ksa1}
\end{figure}

Fig.~\ref{fig:ksa1} shows the baseline KSA schematic for 8-bit operands\footnote{We use a 20-bit KSA in the actual design, but illustrate an 8-bit KSA's schematic here for simplicity.}. First, the KSA computes the sum bits without carry by element-wise XORing of the operands. Next, it keeps combining the generates and propagates in parallel until all the group generates and propagates for each position represents the group from that position until the least significant bit, effectively creating a direct dependence of each carry to the carry-in of the adder. Finally, the adder computes the carry at each position directly from the final group generate, group propagate, and the carry-in and XORs it with the sum without carry to calculate the result.

\begin{figure}
  \centering
    \includegraphics{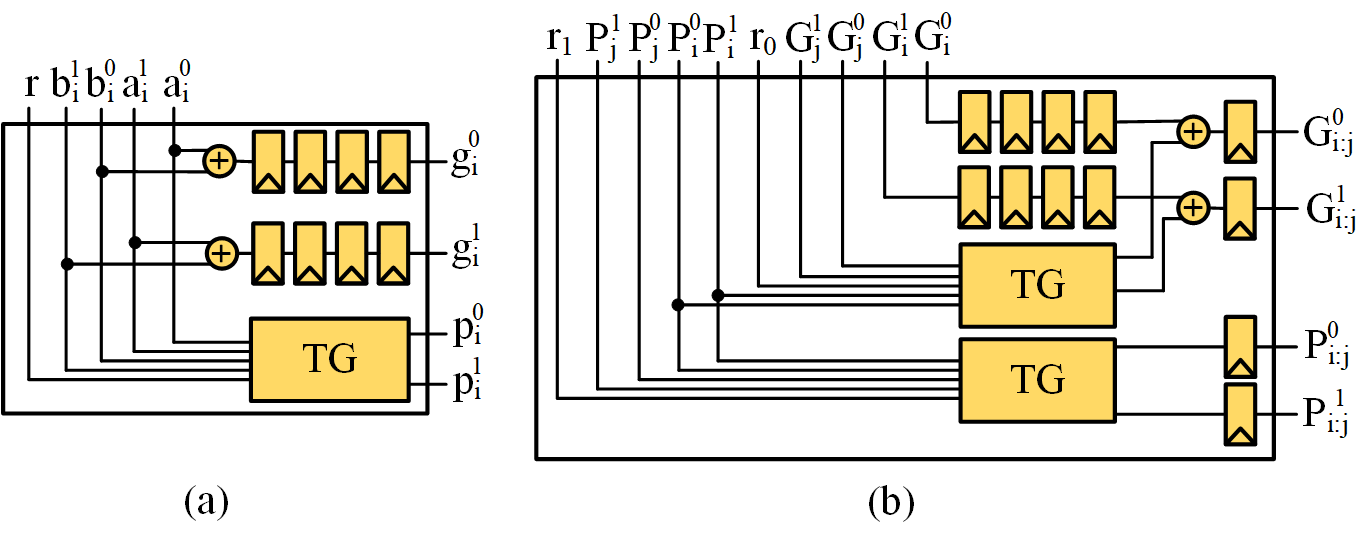}
      \caption{Part (a) describes the masking of the individual generate and propagate blocks. Part (b) shows the masked hardware for the group generate and propagate blocks.}
      \vspace{-1em}
      \label{fig:ksa2}
\end{figure}

We have designed a masked version of the KSA.
Fig.~\ref{fig:ksa2} shows the schematic of this design. 
The adder in the masked design receives two Boolean shares of each operand. 
The function to compute the sum without carry is an XOR operation, which is linear. Therefore, the adder directly and independently computes the Boolean shares of the sum without carry. Similarly it also computes the Boolean shares of the individual propagate bits because they only involve an XOR operation (Equation \ref{eq:prop}). 
The computation of the individual generate bits, however, involve the non-linear AND operation. 
The adder thus replaces the regular AND gates with the synchronised Trichina's AND gate discussed in subsection \ref{ss:tgate} to produce the Boolean shares of the generate bits. 
The produced generate and propagate bits are propagated down the adder tree and combined in each stage using the combination function from Equations \ref{eq:Ggen} and \ref{eq:Gprop}. 
The combination function uses 1 XOR and 2 AND gates. 
The XOR, again, does not require any additional masking, whereas two AND gates are replaced by two synchronised Trichina's AND gates. 
In the final stage, the adder XORs the respective shares of the sum without carry and the generate bits to calculate the Boolean shares of the actual sum.

In the masked KSA architecture the adder's latency is reduced from 100 cycles of masked RCA to 30 cycles. The reduced latency directly helps to reduce the area of the throughput-optimization circuitry that is discussed in Subsection \ref{sec:sched}. The adder latency decides the number of buffers needed to store the concurrent partial summations and the data width of the corresponding demultiplexer and multiplexer blocks. For instance, in the RCA-based design the hardware uses 101 buffers to compute the summations for 101 nodes concurrently because the adder only produces 1 output in 101 cycles.

\subsection{Masking of Activation Function}
\begin{figure}
  \centering
    \includegraphics{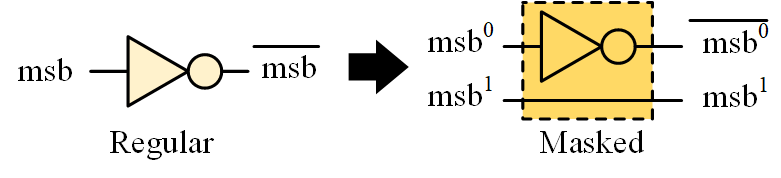}
      \caption{The unmasked activation function (left) is implemented as a NOT gate and the masked version (right) receives two Boolean shares of MSB from masked adder and inverts one of them.}
      \vspace{-1.5em}
      \label{fig:m-actfn}
\end{figure}

The baseline hardware implements the activation function as an inverter as discussed in \ref{sec:actfn}. In the masked version, the MSB output from the adder is a pair of Boolean shares. To perform NOT operation in a masked way, the hardware simply inverts one of the Boolean shares as Fig.~\ref{fig:m-actfn} shows.

\subsection{Masking the Output Layer}

The hardware stores the 10 output layer summations in the form of Boolean shares. To determine the classification result, it needs to find the maximum value among the 10 masked output nodes. Specifically, it needs to \emph{compare} two signed values expressed as Boolean shares. We transform the problem of \emph{masked comparison} to \emph{masked subtraction}. 

\begin{figure}
  \centering
    \includegraphics{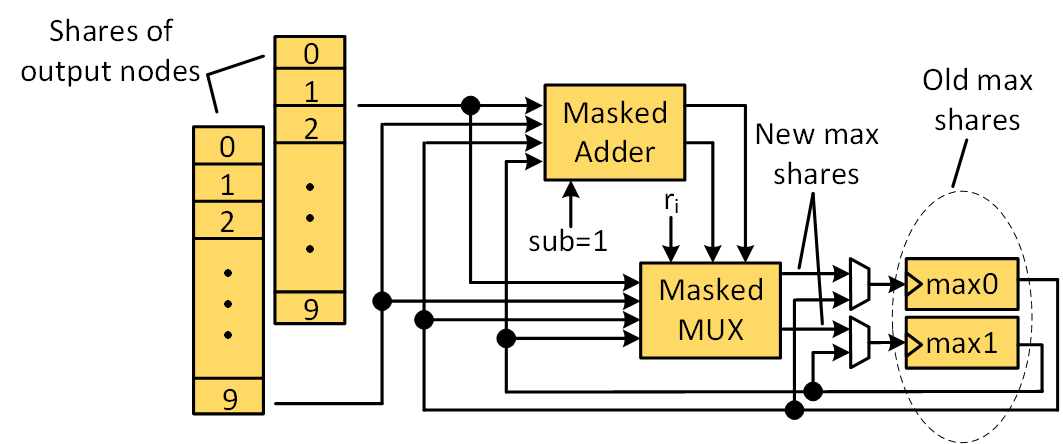}
      \caption{Masking of the Output Layer that uses a masked subtractor and a masked multiplexer to find the node with the maximum confidence score among the 10 output nodes.}
      \vspace{-1em}
      \label{fig:outfn}
\end{figure}

Fig.~\ref{fig:outfn} shows the hardware design of the masked output layer. The hardware subtracts each output node value from the current maximum and swaps the current maximum (old max shares) with the node value (new max shares) if the MSB is 1 using a masked multiplexer. An MSB of 1 signifies that the difference is negative and hence the new sum is greater than the latest max. Instead of building a new masked subtractor, we reuse the existing masked adder to also function as a subtractor through a $sub$ flag, which is set while computing max. In parallel, the hardware uses one more masked multiplexer-based update-circuit to update the Boolean shares of the index corresponding to the current max node (not shown in the figure). This is to prevent known-ciphertext attacks, ciphertext being the classification result in our case. Finally, the Masked Output Logic computes the classification result in the form of (Boolean) shares of the node's index having the maximum confidence score.

Subtraction is essentially adding a number with the 2's complement of another number. 2's complement is computed by taking bitwise 1's complement and adding 1 to it. A bitwise inverse is implemented as an XOR operation with 1 and the addition of 1 is implemented by setting the initial carry-in to 1. The additional XOR, being a linear operator, changes nothing with respect to the masking of the new adder-subtractor circuit.

\subsection{Scheduling of Operations}
\label{sec:sched}
We optimize the scheduling in such a way that the hardware maintains a throughput of 1 addition per cycle. 
The latency of the masked 20-bit RCA adder is 100 cycles. Therefore, the result from the adder will only be available after 101 cycles\footnote{Need an additional cycle for the accumulator register as well.} from the time it samples the inputs. The hardware cannot feed the next input in the sequence until the previous sum is available because of the data dependency between the accumulated sum and the next accumulated sum. This incurs a stall for 101 cycles leading to a total of $784*101=79184$ cycles for each node computation. That is a $784\times$ performance drop over the unmasked implementation with a regular adder.

\begin{figure}
  \centering
    \includegraphics{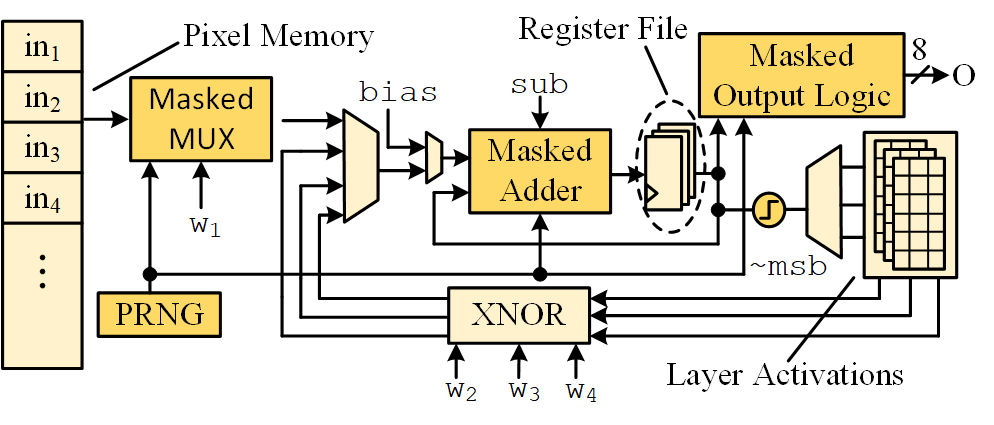}
      \caption{Hardware Design of the Fully Masked Neural Network. The components related to masking are shown in dark yellow. The register file helps in throughput optimization.}
      \label{fig:maskarch}
\end{figure}

 We solve the problem by finding useful work for the adder that is independent of the summation in-flight, during the stalls. We observe that computing the weighted summation of one node is completely independent of the next node's computation. The hardware utilizes this independence to improve the throughput by starting the next node computation while the result for the first node arrives. Similarly, all the nodes up till 101 can be computed upon concurrently using the same adder and achieve the exact same throughput as the baseline design. This comes at the expense of additional registers (see Fig.~\ref{fig:maskarch}\footnote{The register file also has a demultiplexing and multiplexing logic to update and consume the correct accumulated sum in sequence, which is not shown for simplicity.}) for storing 101 summations\footnote{This is why we use 1010 neurons, which is a multiple of 101, in the hidden layers.} plus some control logic but a throughput gain of 784$\times$ (or 1010$\times$ in hidden layers) is worthwhile. The optimization only works if the number of next-layer nodes is greater than, and a multiple of 101. This restricts optimizing the output layer (of 10 nodes) and contributes to the 3.5\% increase in the latency of the masked design.

The masked KSA design extension that we propose has a latency of only 31 cycles. This directly impacts the size of the throughput optimization circuit. Since the adder produces a result in every 32\footnote{Need an additional cycle for the accumulator register.} cycles, the hardware requires a register file of only 32 entries to store the parallel partial summations and accordingly a multiplexer and demultiplexer of width 32. Hence we reduce the area cost of the throughput optimization circuitry by 3$\times$ which is coherent with the synthesis results shown in Table \ref{tab:maskBlk}.

\section{Shuffling the Neural Network Computations}
\label{sec:shuffle}
Shuffling is a commonly used technique to defend against side-channel attacks~\cite{shuffle-survey}. The premise of shuffling is to introduce noise in the temporal domain by either randomizing the order of independent operations and/or by adding random delays within the algorithm. 
In DPA attacks, the adversary tries to correlate its hypotheses with the side-channel measurements at a fixed point of interest over multiple measurements assuming that the targeted operation happens \emph{at the same time in each trace}. 
Since in a shuffled implementation the targeted operation will happen at the same time instance in a small set of traces, the number of traces needed for a successful attack increases. 
Prior work by Mangard \textit{et al.} demonstrates that the number of traces required for a successful attack is $p^2$ if the probability of occurrence of the operation of attack is $1/p$~\cite{mangard2008power}. 
Therefore, shuffling would increase the number of measurements needed to run a successful side-channel attack by $p^2$. 
Shuffling improves the side-channel resilience in general for all orders due to the temporal perturbation introduced. However, we note that our proposed masking scheme already provides first-order side-channel security and thus, the adversary will naturally try a second-order attack as the next step. Therefore, shuffling makes it more difficult to conduct a second-order side-channel attack on our first-order masked design.

Earlier studies have adopted shuffling as a countermeasure against physical side-channel attacks on both software and hardware implementations of cryptographic primitives \cite{shuffle-survey}. Depending on the performance and area budget, the shuffle is implemented either as random permutation (RP) or as a random starting index (RSI)~\cite{rsi1}. RP-based shuffling typically uses the Fisher-Yates shuffle algorithm~\footnote{Also known as the Knuth algorithm} that generates unbiased permutations of a given sequence with $n$ elements leading to $n!$ possible permutations. 
In RSI, the original sequence is cyclically rotated by a random chosen amount, which is the starting index, leading to $n$ possible permutations. 
We identify that the computation of the activation value of each node in a layer is independent of other nodes. This allows the hardware to shuffle the order of computation of activation values, which is by default done sequentially in the unshuffled implementation. The RSI-based shuffle is almost free in hardware because only the starting read address of the memory storing the parameters needs to be initialized with a random index. This makes RSI a very good candidate for a low area-budget design. 
Given the area overheads and constant-time enforcement challenges of Fisher-Yates shuffle, we have opted for applying the RSI in this work. 

RSI-based shuffling enables the number of possible permutations to be equal to the number of nodes in a hidden layer, i.e., 1024. Based on the analysis by Mangard \textit{et al.} \cite{mangard2008power}, the number of traces for a successful attack increases by $1024^2\times$, i.e., about six orders-of-magnitude. The large number of independent operations in neural networks helps to provide a significantly higher side-channel security using RSI, compared to that in block ciphers, which is typically in two orders-of-magnitude due to fewer independent operations, for instance, the number of SBox computations. Implementing RP-based shuffling offers a better security but has significant area costs due to the following reasons. Every step of the Fisher-Yates shuffle requires generating a uniform random number within a dynamically changing range. However, PRNGs typically generate random numbers only within ranges that are powers of 2. One way to confine the generated random number within the required range is to perform a modulo but that leads to biased permutations~\cite{non-const-rng}. This is a well-studied problem and one way to solve it is by discarding certain random numbers that create the bias (a.k.a. rejection sampling) that makes the algorithm variable-time, and possibly introduce timing-based side-channel vulnerabilities. The rejection sampler and modulo operation with non-powers of 2 incur significant area overheads when implemented on hardware. 

In the shuffled neural network implementation, the hardware generates a random index to start the node computations in a layer. 
The hardware then evaluates all the nodes sequentially from the randomly chosen node until the last node and then the remaining nodes from the first node until the starting node. 
The hardware can further shuffle the reads of the input and hidden layer nodes during weighted summation because because addition is commutative. However, we do not chose to implement that shuffling in this work because the number of input layer nodes is 784, which is not a power of 2 and hence requires a variable time random number generator for an unbiased generation of numbers between 0 and 783~\cite{non-const-rng}.

%% file: 5.Results.tex
\section{Results} 
In this section, we describe the hardware setup used to implement the neural network and capture power measurements, the leakage assessment methodology that we follow to evaluate the security of the proposed design, and the hardware implementation results.

\subsection{Hardware Setup}
We use Verilog for the front-end design and Xilinx ISE 14.7 for the back-end. We use the DONT\_TOUCH synthesis attribute and disable the tool options like LUT combining, register reordering, etc. to prevent any optimization in the masked components. These are standard practices in implementing masking schemes on FPGAs.

Our side-channel evaluation platform is the SAKURA-G FPGA board \cite{sakurag}. It hosts Xilinx Spartan-6 (XC6SLX75-2CSG484C) as the main FPGA that executes the neural network inference. An on-board amplifier amplifies the voltage drop across a 1$\Omega$ shunt resistor on the power supply line. We use Picoscope 3206D \cite{picoscope} as the oscilloscope to capture the measurements from the board. We conduct different experiments with varying design and sampling frequencies to shorten the time of acquisition and evaluation and to also validate a sound measurement setup. 
\begin{enumerate}
    \item We use the design and sampling frequencies of 24MHz and 125MHz, respectively, for the experiments presented in the conference version. Since we evaluate the complete inference time window for \texttt{Variant-I}, we had to increase the design frequency to shorten the inference latency and accelerate the validation process.
    \item We use the design and sampling frequencies of 1.5MHz and 500MHz, respectively, to evaluate the synchronized Trichina's AND gate, \texttt{Variant-II}, and \texttt{Variant-III}. Synchronized Trichina's AND gate is a small hardware primitive with a latency of only 4 clock cycles, and we only capture small time windows to evaluate variants \texttt{II} and \texttt{III}. Therefore, we can decrease the design frequency without any significant time overheads in the leakage evaluation process.
\end{enumerate} 
We discuss more on how the high latency of the design poses challenges in evaluation in Section~\ref{measDifficulty}.

We use Riscure's Inspector SCA \cite{inspector} software to communicate with the board and initiate a capture on the oscilloscope. By default, the Inspector software does not support SAKURA-G board communication. Hence, we develop our own custom software modules to automate the setup. The modules implement the communication protocol and perform the register reads/writes on the FPGA to start the inference and capture the result.

\subsection{Leakage Evaluation}
We perform the leakage assessment of the proposed design using the non-specific fixed vs random t-tests, which is a common and generic way to assess the side-channel vulnerability in a given implementation \cite{becker2013test}. Each input is an image consisting of 784 8-bit unsigned pixels. Since we run a non-specific TVLA test, our setup first selects an image to be used as the fixed image. Then, it sends fixed and random images to the device in a randomly interspersed fashion as prescribed by Schneider et al. in a prior work~\mbox{\cite{schneider2016leakage}}.
A t-score within the threshold range of $\pm$4.5 implies that the power traces do not leak any information about the data being processed with up to 99.99\% confidence. The measurement and evaluation is quite tedious and we refer the reader to Section \ref{measDifficulty} for further details. We demonstrate the security up to 2M traces, which is much greater than the first-order security of the currently best-known defense \cite{dubey2019maskednet}. 

Pseudo Random Number Generators (PRNG) produce the fresh, random masks required for masking. 
We choose TRIVIUM~\cite{prng} as the PRNG, which is a hardware implementation friendly PRNG specified as an International Standard under ISO/IEC 29192-3, but any cryptographically-secure PRNG can be employed.
TRIVIUM generates 2$^{64}$ bits of output from an 80-bit key; hence, the PRNG has to be re-seeded before the output space is exhausted. Note that this randomness generation is not a unique problem for masking NN and do exist in masking crypto circuits; hence, we adopt their approach.

Next we present the security evaluation for three design variants. {\tt Variant-I} is the design that uses only the masked RCA, which is taken from the shorter, conference version of this article~\cite{bomanet}. {\tt Variant-II} is the masked KSA-based design that we propose as part of our extensions. {\tt Variant-III} incorporates the shuffling of node computations as well along with the masking of the KSA, which is also a part of the extensions.

\begin{figure}
  \centering
    \includegraphics[width=0.8\textwidth]{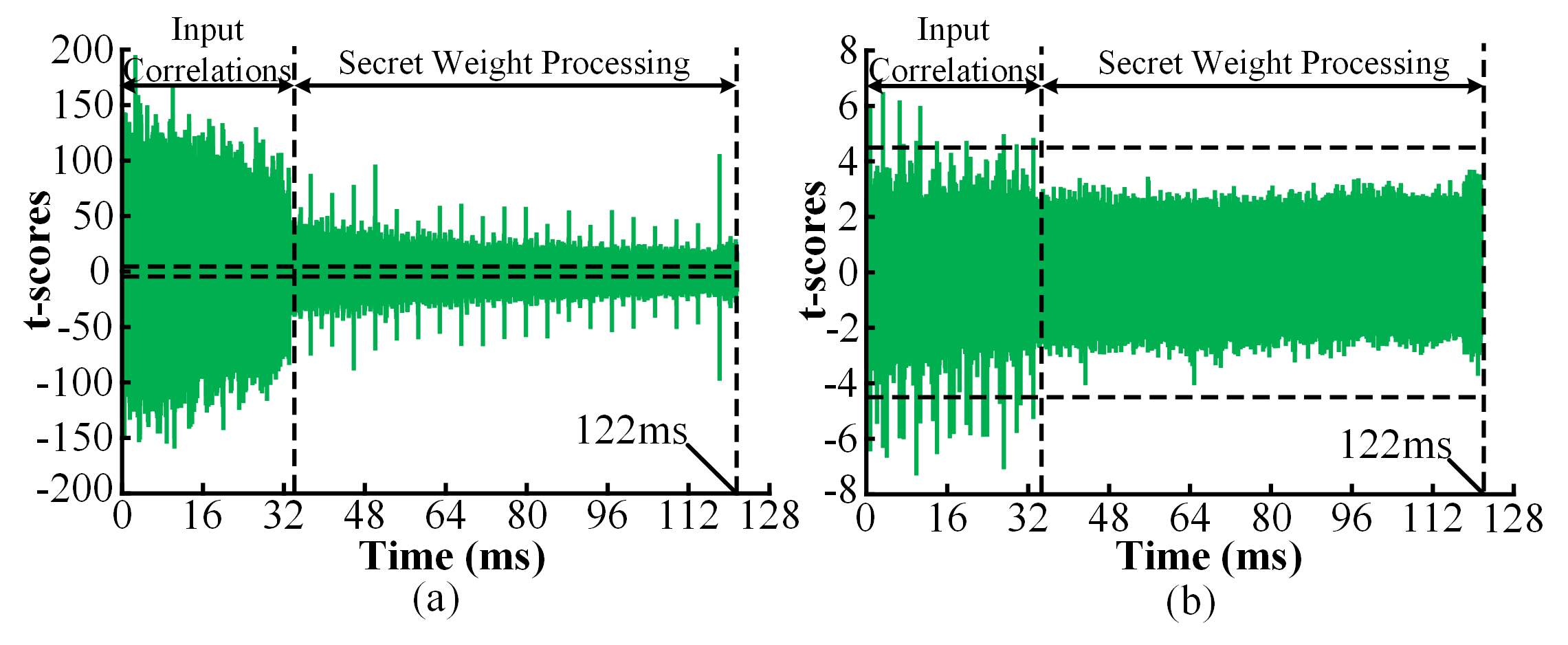}
    \caption{TVLA results of the unmasked (a) and masked (b) RCA-based implementation. The results clearly show that the unmasked design is insecure, whereas the masked design is secure with 99.99\% confidence (t-scores always below ±4.5).}
    \vspace{-1em}
    \label{fig:tval_rca}
\end{figure}

\subsubsection{First-order tests on {\tt Variant-I}}
First, we conduct a first-order t-test on the design with PRNGs disabled, which is equivalent to an unmasked (baseline/unprotected) design. Fig.~\ref{fig:tval_rca} (a) shows the result of this experiment where we clearly observe significant leakages throughout the execution since the t-scores are greater than the standard threshold of $\pm$4.5. Next, we conduct the same test, but with PRNGs enabled, which is the masked design. Fig.~\ref{fig:tval_rca} (b) shows the results for this case, where we observe that the t-scores never cross the threshold of $\pm$4.5 except a small portion in the start of the plot. 

The leakages in the initial portion of the plot are due to input correlations during the input layer computations. The hardware loads the input pixel after every 101 cycles and feeds it to the masked multiplexer. However, the secret variable is the \emph{weight} and not the input pixel, which is never exposed because the masked multiplexer randomises the output using a fresh, random mask. 

\subsubsection{High Precision First and Second-order tests on {\tt Variant-I}}
\begin{figure}
  \centering
    \includegraphics[width=0.7\textwidth]{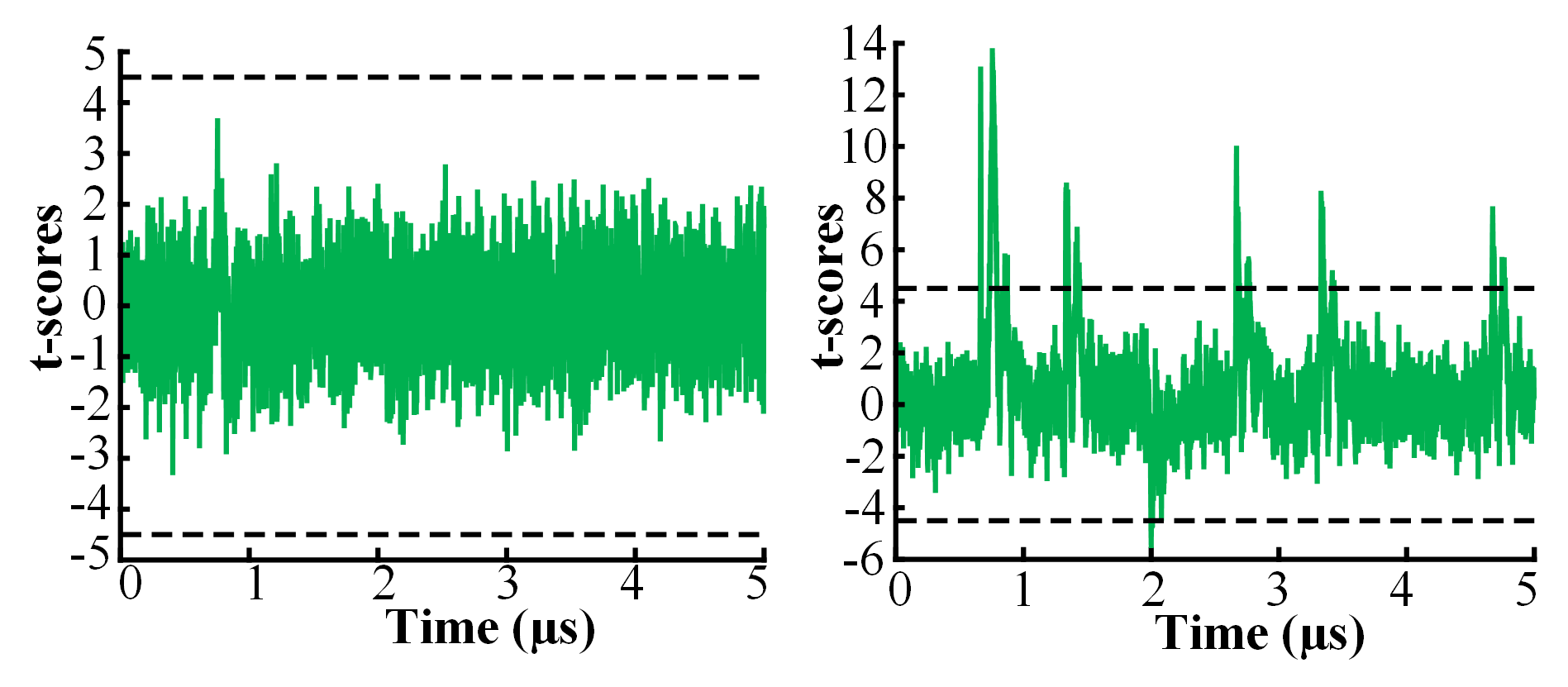}
    \caption{First-order (left) and second-order (right) t-tests on Trichina's AND gate at a low design frequency of 1.5MHz and sampling frequency of 500MHz.}
    \vspace{-1em}
    \label{fig:tval_hp}
\end{figure}

We performed univariate second-order t-test on the fully masked design \cite{schneider2016leakage} but 1M traces were not sufficient to reveal the leakages. Due to the extremely lengthy measurement and evaluation times it was infeasible to continue the test for more traces. Therefore, we perform first and second-order evaluation on the isolated synchronized Trichina's AND gate, which is one of the main building blocks of the design. We reduce the design frequency to 1.5MHz to increase the accuracy of the measurement and prevent any clock cycle aliasing. The signal-to-noise ratio (SNR) for a single gate was not sufficient to see leakage even at 10M traces, hence we amplify the SNR by instantiating 32 independent instances of the Trichina's AND gate in the design, driven by the same inputs. We present the results for this experiment in Fig.~\ref{fig:tval_hp} that shows no leakage in the first-order t-test but significant leakages in the second-order t-tests for 500k traces. Thus, the successful second-order t-test validates the correctness of our measurement setup and the first-order masking implementation.

\subsubsection{First-order tests on {\tt Variant-II} and \texttt{Variant-III}}
\label{sec:sca-ksa}
\begin{figure}
  \centering
    \includegraphics[width=1.0\textwidth]{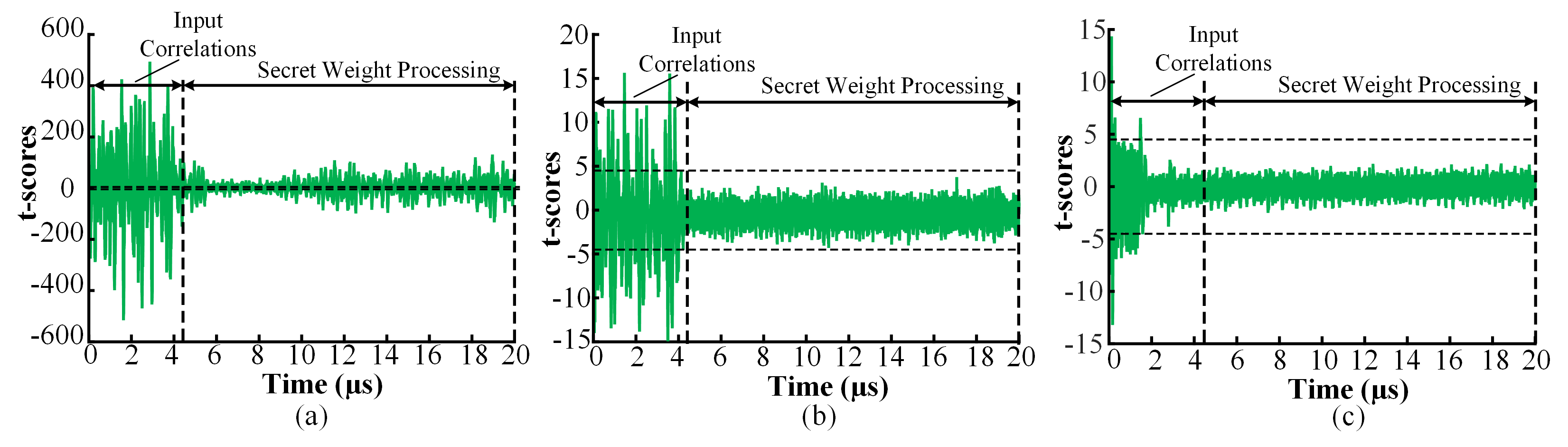}
    \caption{First-order t-test results of the KSA-based design with PRNGs off (a), PRNGs on (b), and PRNGs on coupled with RSI shuffling of the node computations (c). (a) shows clear leakages throughout the execution, (b) only shows leakages during the input layer computations, and (c) shows a significantly lower leakage during the input layer computations as well due to the shuffle.}
    \label{fig:tval_ksa}
\end{figure}

\begin{figure}
  \centering
    \includegraphics[width=1.0\textwidth]{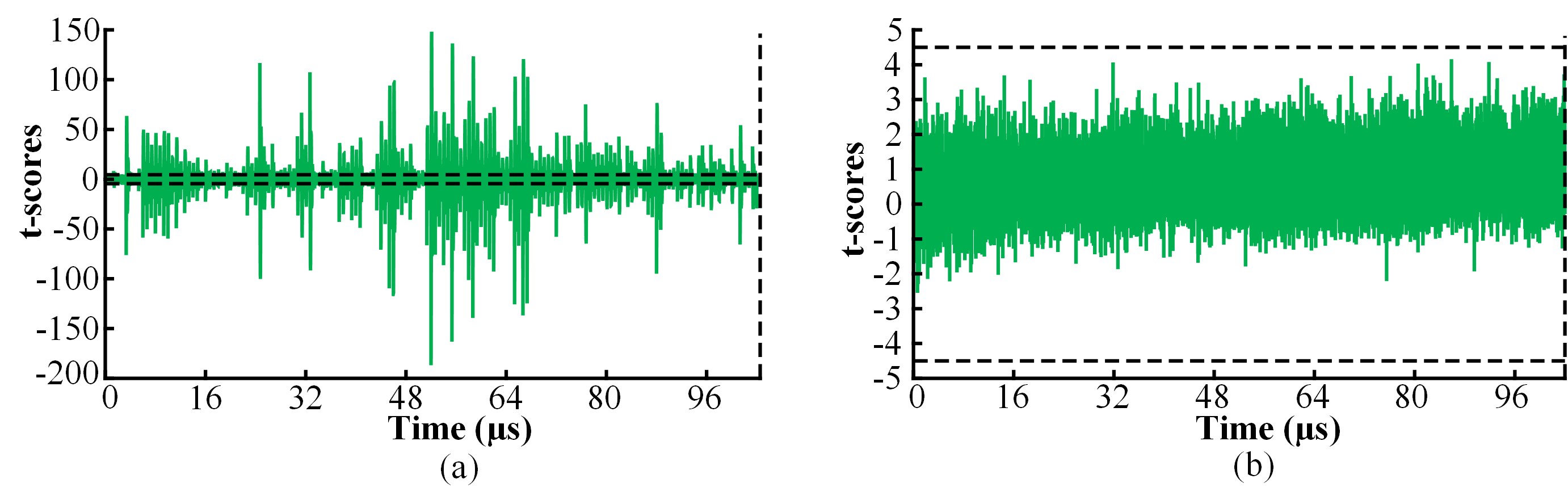}
    \caption{First-order t-test results of the output layer computations in the KSA-based design with PRNGs off (a), PRNGs on (b). (a) shows clear leakages throughout the execution, whereas (b) shows that the t-scores stay confined to the threshold of $\pm$4.5.}
    \label{fig:tval_ol}
\end{figure}

The tests involve (i) increasing the sampling frequency of the oscilloscope to capture potential glitch-caused leakages more effectively and (ii) evaluating the canonical layers' computation rather than the entire scheme to achieve a practical validation. Note that this is a standard practice in side-channel analysis---e.g., only the first few exponentiations are analyzed in a 1024-bit RSA scheme~\cite{DLRSA}. 
We present our evaluations for two short time windows out of the overall inference, which involve unique computations, for {\tt Variant-II}. The setup captures the first time window when hardware transitions from the input layer computations to the first hidden layer computations (Fig.~\ref{fig:tval_ksa}). The second time window is the complete evaluation of the output layer which involves the computation of maximum confidence using the newly proposed masked KSA (Fig.~\ref{fig:tval_ol}). 
Fig.~\ref{fig:tval_ksa} (a) shows that the first-order t-test with PRNGs disabled clearly leaks during the computations of both the input and hidden layer. However, in the masked design i.e., with PRNGs switched on, the t-scores never cross the threshold of $\pm$4.5 during the hidden layer computations. The results for \texttt{Variant-III} are similar with an additional decrease in the t-scores because of increased temporal noise due to shuffling. Fig.~\ref{fig:tval_ksa} (c) shows the TVLA results for \texttt{Variant-III}. 

Fig.~\ref{fig:tval_ol} shows the TVLA results for the entire output layer computation with PRNGs off and PRNGs on. The results again demonstrate that the unmasked design leaks significantly, whereas the masked design's leakage is within the standard threshold of $\pm$4.5 throughout the computation of the masked classification result.

\subsubsection{Evaluation of Frequency Traces}
\begin{figure}
  \centering
    \includegraphics[width=1.0\textwidth]{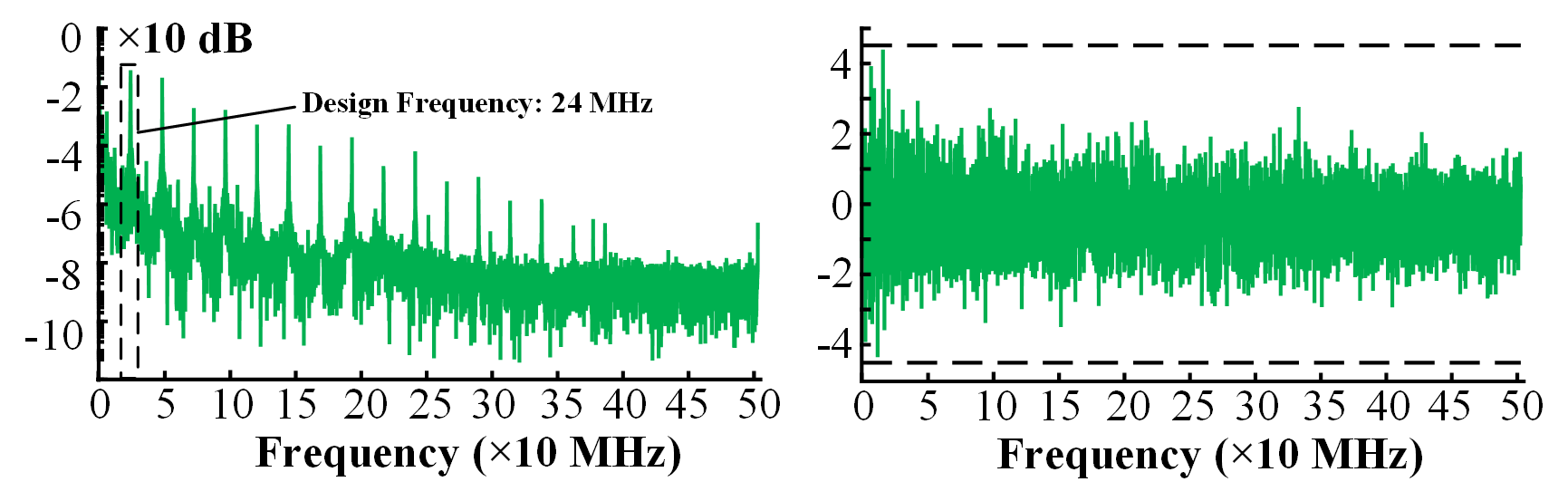}
    \caption{Average frequency spectrum of masked \texttt{Variant-III} (left) and its TVLA result (right). The TVLA plot shows that there is no significant leakage in the frequency spectrum of the design when masked.}
    \label{fig:freqtvla}
\end{figure}

Converting the power traces to frequency domain is one potential way to nullify the effect of shuffling because shuffling only introduces noise in the time domain. Thus, the adversary may conduct a side-channel attack on the frequency domain traces to extract the secret information. However, masking the design reduces the leakages in the frequency domain as well and resists a first-order DPA on the frequency traces. We conduct a TVLA test on the frequency domain traces of \texttt{Variant-III} to quantify the side-channel information in the frequency domain. Prior works have also used frequency domain TVLA analysis to quantify leakages in the frequency domain~\cite{singh2018improved,lei2017frequency}. Fig.~\ref{fig:freqtvla} shows the average frequency trace and the first-order TVLA results with 2M traces. We observe that the t-scores are within the threshold for the frequency traces implying that the design does not leak information in the frequency domain as well.

\subsection{Masking Overheads}
\label{sec:maskov}
Table~\ref{tab:maskOv} summarizes our implementation results for the baseline unmasked design and all the proposed variants. We observe that the {\tt Variant-I} incurs an area overhead of 5.4$\times$ and 6.8$\times$ in the LUTs and FFs compared to the baseline. {\tt Variant-II}, however, reduces these overheads to 4.7$\times$ and 6.6$\times$ respectively, for the LUTs and FFs. The primary reason for the reduction is the use of KSA instead of RSA that reduces the latency of the adder and in turn also reduces the hardware cost for throughput optimization. {\tt Variant-III} which incorporates RSI-based shuffling along with masking shows a negligible increase of 16 LUTs and 10 FFs compared to {\tt Variant-II}.

We also present a block-wise area distribution of the RCA and the KSA-based designs in Table \ref{tab:maskBlk} to better understand the reductions. The fourth column shows the increase or decrease in the area of a particular block in {\tt Variant-III} compared to {\tt Variant-I}. As expected, we notice an increase in the area of the KSA compared to the RSA due to more combinational logic like the generate and propagate circuitry. The number of PRNGs required is also higher in {\tt Variant-III} due to an increase in the number of AND gates in the KSA compared to RSA that demand additional fresh random masks. However, we can observe the main contributor of the overall area in {\tt Variant-I} is the Throughput Optimization circuitry and we reduce the cost for this block by roughly 3$\times$ in {\tt Variant-III}. Since the number of nodes in the hidden layers of {\tt Variant-III} is 1023 instead of 1010 in {\tt Variant-I}, we observe slight increase in the read-only memories (ROM) that store the parameters and the read-write memories (RWM) that store the node values. The latency in both the implementations increases only by a mere 1.03$\times$ compared to baseline unmasked design, primarily due to absence of the throughput optimization in the output layer. 

Next, we compare the area-delay product (ADP) of the proposed Variants with the only work published on side-channel countermeasures for neural networks called MaskedNet~\cite{dubey2019maskednet}. Here, we define area as the sum of the number of LUTs and FFs, and delay as the latency in number of cycles. 
We emphasize that a direct comparison between these approaches in unfair because MaskedNET inherently gains an advantage due to its parallelized design (compared to BoMaNET's sequentialized datapath). 
Indeed, the comparison of the baseline (unprotected) design of parallel vs. sequential hardware reveals a 68$\times$ area-delay product advantage for the parallel hardware. 
The ADP increase in the masked versions i.e., between {\tt Variant-II} and protected MaskedNet design, is 80$\times$, which is a relative increase of only 1.18$\times$. Our improved design {\tt Variant-III} reduces the relative increase in ADP to 1.08$\times$. 

Similar overheads were observed in previous works on bit-sliced masked AES implementations~\cite{maskOverhead}.
\begin{table}[t!]
  \centering
  \caption{Area (LUT/FF/BRAM) and Latency (in cycles) Comparison of the Unmasked and Masked Implementations.}
  \begin{tabular}{ |c|c|c|c| } 
     \hline
     \textbf{Variant} & \textbf{Area} & \textbf{Latency} & \textbf{Change}\\ 
     \hline
     Unmasked & 1833/1125/163 & \num{2.85e6} & NA\\
     \hline
     {\tt Variant-I} & 9818/7709/163 & \num{2.94e6} & 5.4$\times$ / 6.8$\times$ / 1$\times$\\ 
     \hline
     {\tt Variant-II} & 8724/7483/165 & \num{2.94e6} & 4.7$\times$ / 6.6$\times$ \ 1$\times$\\ 
     \hline
     {\tt Variant-III} & 8740/7493/165 & \num{2.94e6} & 4.7$\times$ / 6.6$\times$ \ 1$\times$\\ 
     \hline
  \end{tabular}
  \label{tab:maskOv}
\end{table}

\begin{table}[t!]
    \begin{threeparttable}
    \centering
    \caption{Block-level Area Distribution of the Unmasked and Masked Implementations (LUT/FF/BRAM).}
    \begin{tabular}{ |c|c|c|c| } 
        \hline
        \textbf{Design Blocks} & {\tt \textbf{Variant-I}} & {\tt \textbf{Variant-III}} & \textbf{Fraction (\%)}\\ 
        \hline
        Adder & 929/1135/0 & 1572/2074/0 & (+) 1.7/1.8/\\ 
        \hline
        PRNGs & 1188/1314/0 & 3112/2916/0 & (+) 2.5/2.4/\\ 
        \hline
        Output Layer & 54/82/0 & 50/82/0 & (-) 1.08/1/-\\ 
        \hline
        Throughput & 5486/4126/0 & 1792/1338/0 & (-) 3.1/3.1/-\\
        Optimization & & &\\
        \hline
        Controller & 1471/43/0 & 1513/56/0 & (+) 1.02/1.3/-\\ 
        \hline
        ROMs & 1009/670/159 & 1027/677/161 & (+)1.01/1.01/1\\ 
        \hline
        RWMs & 0/0/4 & 0/0/4 & -/-/1\\
        \hline
    \end{tabular}
    \label{tab:maskBlk}
    \begin{tablenotes}
        \small
        \item "-" denotes no change in the area of the unmasked and masked design.
        \item "(+)/(-)" in Fraction column denotes an increase/decrease in the area of {\tt Variant-III} compared to {\tt Variant-I}.
    \end{tablenotes}
    \end{threeparttable}
\end{table}

%% file: 6.Disc.tex
\section{Discussions}

\subsection{Proof-of-Concept vs. Optimizations}
Our solutions utilize simple yet effective techniques to mask an inference engine. But certainly, there is scope for improvement both in terms of the hardware design and the security countermeasures. 
In this section, we discuss possible optimizations/extensions of our work and alternate approaches taken in the field of privacy for ML.
\subsubsection{Design Optimizations}
\label{sec:luttrichina}
\begin{figure}
  \centering
    \includegraphics[width=0.8\textwidth]{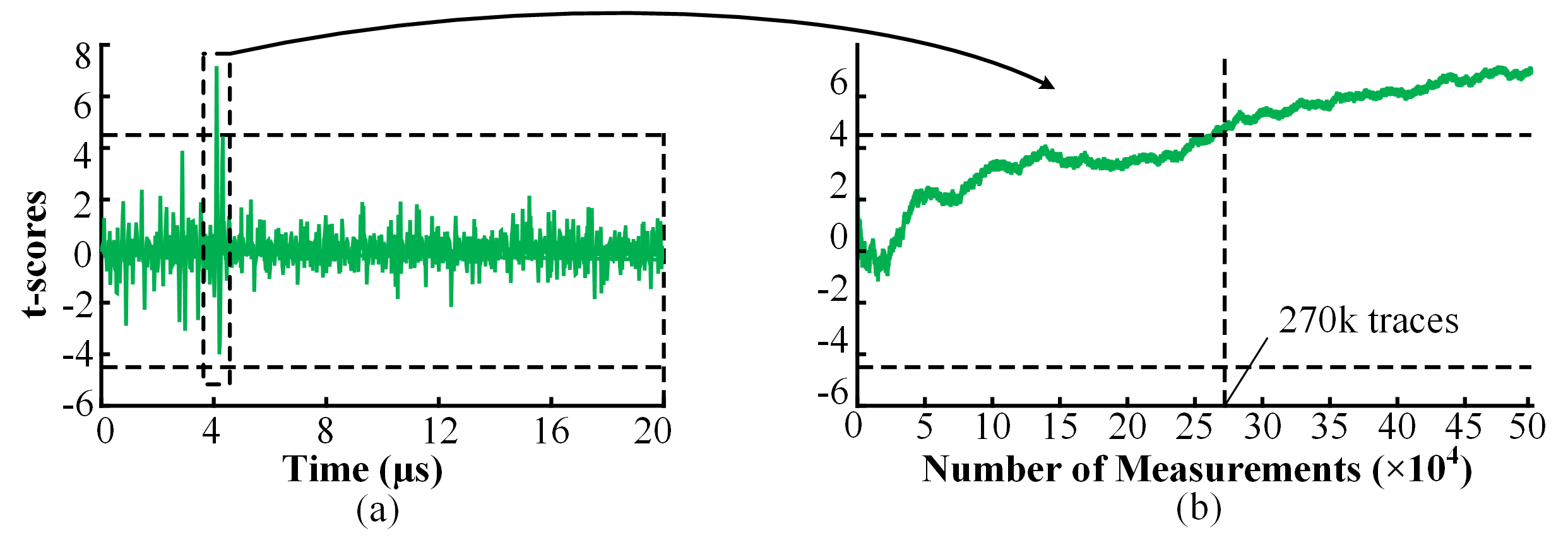}
    \caption{TVLA results of LUT-based implementation of Trichina's AND gate.}
    \label{fig:tval_tglut}
\end{figure}

We implement the Trichina's AND gate as a fully-pipelined gate-level masked structure to resist leakages from the glitches. However, since it only utilizes 5 inputs and our target platform is an FPGA, it can also be implemented directly as a 5-input masked look-up table (LUT). Such a design will significantly simplify the design, and reduce the latency of the masked AND operation and area costs due to pipeline registers. Reducing the latency of the adder will also decrease the area cost of the throughput optimization, which is what we observe when we switch from RCA to KSA as well. Therefore, we conducted TVLA tests on the LUT-based design as well to quantify the area costs and the security loss. Fig.~\ref{fig:tval_tglut} shows the first-order t-test results on the masked full adder implemented using LUT-based Trichina's AND gates. We observe that the design starts leaking around 270k traces and area of the overall design decreases by 3$\times$ and 2.7$\times$ in the number of FFs and LUTs respectively. The leakage is because of the glitches that happen inside the LUTs of the FPGA due to different arrival times of the inputs. These glitches can temporarily reconstruct the secret intermediates inside the LUT.

Prior literature discusses other techniques to resist glitches. For instance, the prior work on TI-based secure versions of RCA and KSA can be extended for this work~\cite{boma-adder15}. Another potential optimization is to selectively allow only those glitches to happen that do not reconstruct the secret, and register every other combinational output. This idea has been well explored in the work by Gross \textit{et al.} on Domain Oriented Masking \cite{gross2016domain}. Another prior work tried to prevent glitches inside the LUTs by disabling the toggles in the first half of the clock cycle using an enable signal as an input to the LUT  \cite{manualGlitch}. However, it required implementing the AES SBox by manually hard instantiating each LUT making it somewhat tricky to be extended for larger designs like a neural network. 

The ultimate goal of a side-channel defense is to make the design as secure as the primary-channel attack, which is theoretical cryptanalysis.  For cryptography, this means making the design secure against an astronomical amount of tests (e.g., 2$^{128}$ for AES).  But this is not necessarily the case for machine learning.  Recent works shows that a high-fidelity, cryptanalytic model extraction succeeds with a few million tests on the types of networks that are similar to those we evaluate in this work~\cite{carlini2020cryptanalytic}. Therefore, it would be sufficient for the side-channel defense to be secure against a test using a few million traces because the design will be broken anyhow with that many tests through a cryptanalytic attack. The lowered security requirement of ML (compared to cryptographic primitives) would enable designers to opt for a solution that use a glitch-vulnerable Trichina's AND gate.

\subsubsection{Limitations}
We reduce glitch-related vulnerabilities using registers at each stage, which is a low-cost, practical solution. Other works have proposed stronger countermeasures at the cost of higher performance and area overheads~\cite{ICICS:NikRecRij06,gross2016domain}.
The quest for stronger and more efficient countermeasures is never-ending; masking of AES is still being explored, even 20 years after the initial work~\cite{CHES:AkkGir01}, due to the advent of more efficient or secure masking schemes \cite{d+1shares} and more potent attacks~\cite{moos2017static,TKL05}. 

Our solution is first-order secure but there is scope for construction of higher-order masked versions. However, higher-order security is a delicate task; Moos \textit{et al.} recently showed that a straightforward extension of masking schemes to higher-order suffers from both local and compositional flaws \cite{TCHES:MMSS19} and masking extensions were proposed in another recent work \cite{cassiershardware}. This is the first work on fully-masked neural networks and we foresee follow ups as we have experienced in the cryptographic research of AES masking, even after 20 years of intensive study.

We consider fault attacks out of scope for this work but it would be interesting to explore defenses for such attacks. Numerous countermeasures have been proposed to defend cryptographic implementations against fault attacks~\cite{de2018m}. Therefore, one promising direction could be to explore how to extend those ideas in the context of machine learning algorithms. Another potential research direction is to explore efficient side-channel attacks to extract the model weights and biases, which might become tedious for deeper networks. For example, if the neural network layers are similar, adversary may create a template of power consumption for the initial layer computations and use it to attack the subsequent layers.

\subsection{Measurement Challenges}
\label{measDifficulty}
We faced some unique challenges that are not generally seen with the symmetric-key cryptographic evaluations. Inference becomes a lengthy operation, especially for an area-optimized design---the inference latency of our design is roughly 3 million cycles. For a design frequency of 24MHz, the execution time translates to 122ms per inference. If the oscilloscope samples at 125MHz the number of sample points to be captured per power trace is equal to 15 million. This significantly slows down the capturing of power traces. In our case, capturing 2 million power traces took one week, which means capturing 100 million traces as AES evaluation~\cite{d+1shares} will take roughly a year. Performing TVLA on such large traces (~28TB, in our case) also takes a significant amount of time: it took 3 days to get one t-score plot during our evaluations on a high-end PC\footnote{Intel Core i9-9900K, 64GB RAM.}. One possibility to avoid this problem is looking at a small subset of representative traces of the computation~\cite{DLRSA}. We conduct a comprehensive evaluation of our design in one variant and an evaluation on a smaller subset in the other variants. 

\subsection{Theoretical vs Side-Channel Attacks}
Theoretical model extraction by training a proxy model on a synthetically generated dataset using the predictions from the unknown victim model is an active area of research \cite{jagielski2019high,carlini2020cryptanalytic}. These attacks mostly assume a black-box access to the model and successfully extract the model parameters after a certain number of queries. This number ranges typically in the order of $2^{20}$ \cite{carlini2020cryptanalytic}. By contrast, physical side-channel attacks only require a few thousand queries to successfully steal all the parameters \cite{dubey2019maskednet}. This is partly due to fact that physical side-channel attacks can extract information about intermediate computations even in a black-box setting. Physical side-channel attacks also do not require the generation of the synthetic dataset, unlike most theoretical attacks.

\subsection{Orthogonal ML Defences}
There has been some work on defending the ML models against stealing of inputs and parameters using other techniques like Homomorphic Encryption (HE) and Secure Multi-Party Computation (SMPC) \cite{cryptonets16,secureml17,xonn19}, Watermarking \cite{rouhani2018deepsigns,adi2018turning}, and Trusted Execution Engines (TEE) \cite{preventingnn18,tramer2018slalom,mlcapsule18}. We refer the interested readers to survey papers published on ML privacy for a more exhaustive list~\cite{mlNIST,isakov2019survey,trust-ai,ml-survey2}. Purely HE-based or HE+SMPC solutions initially incurred a large overhead (30000$\times$ in CryptoNet \cite{cryptonets16} and 500$\times$ in SecureML \cite{secureml17}), but follow-up work made them more efficient (13$\times$ in XONN\cite{xonn19}). We propose masking, which is an extension of SMPC on hardware and we believe that it is a promising direction for ML side-channel defenses as it has been on cryptographic applications. Watermarking techniques are punitive methods that cannot prevent physical side-channel attacks. TEEs are subject to ever-evolving microarchitectural attacks and typically are not available in edge/IoT nodes.

\subsection{Extensions to More Complicated Networks}
We believe that some of the solutions we propose can be used to mask other types of neural networks as well, and are not just confined to the specific MLP architecture we target in this work. The masked adder can be used to mask the convolution operation in Convolutional Neural Networks. 
Our masked output layer design can be used to construct a masked MaxPool hardware that can compute the maximum element in a given pooling window in masked fashion. MaxPool is typically implemented using  OR gates if the network is binarized because MaxPool becomes equivalent to an OR operation. Our proposed synchronized Trichina’s AND gate design can be used to mask the OR operations too.